\documentclass[fleqn,usenatbib]{mnras}

\usepackage{newtxtext,newtxmath}

\usepackage[T1]{fontenc}

\DeclareRobustCommand{\VAN}[3]{#2}
\let\VANthebibliography\thebibliography
\def\thebibliography{\DeclareRobustCommand{\VAN}[3]{##3}\VANthebibliography}

\usepackage{graphicx}	
\usepackage{amsmath}	
\usepackage{longtable}
\usepackage{pdflscape}

\def\xmm {\emph{XMM--Newton}}
\def\cxo {\emph{Chandra}}
\def\nustar {\emph{NuSTAR}}

\def\swift {\emph{Swift}}
\def\nicer {\emph{NICER}}

\def\srclong{SGR\,J1935+2154}
\def\src{SGR\,J1935}

\def\flux {\mbox{erg cm$^{-2}$ s$^{-1}$}}
\def\lum {\mbox{erg s$^{-1}$}}
\def\nh {$N_{\rm H}$}

\def\cmdue {\rm \ cm$^{-2}$}
\def\ss {\mbox{s\,s$^{-1}$}}
\def\chisq {$\chi ^{2}$}

\def\deg{\mbox{$^{\circ}$}}

\def\wes{CXOU\,J1647$-$4552}

\title[The 2020 X-ray outburst of \src]{The first 7 months of the 2020 X-ray outburst of the magnetar \srclong}

\author[A. Borghese et al.]{
A. Borghese$^{1,2}$\thanks{E-mail: borghese@ice.csic.es}, F. Coti Zelati$^{1,2}$, G.L. Israel$^{3}$, M. Pilia$^{4}$, M. Burgay$^{4}$, M. Trudu$^{4}$, S. Zane$^{5}$, R. Turolla$^{5,6}$,
\newauthor
N. Rea$^{1,2}$, P. Esposito$^{7,8}$, S. Mereghetti$^{8}$, A. Tiengo$^{7,8,9}$, A. Possenti$^{4,10}$ 
\\
$^{1}$Institute of Space Sciences (ICE, CSIC), Campus UAB, Carrer de Can Magrans s/n, E-08193, Barcelona, Spain\\
$^{2}$Institut d'Estudis Espacials de Catalunya (IEEC), Carrer Gran Capit\`a 2--4, E-08034 Barcelona, Spain\\
$^{3}$INAF--Osservatorio Astronomico di Roma, via Frascati 33, 00078 Monteporzio Catone, Italy \\
$^{4}$INAF--Osservatorio Astronomico di Cagliari, Via della Scienza 5, I-09047 Selargius, Italy\\ 
$^{5}$Mullard Space Science Laboratory, University College London, Holmbury St. Mary, Dorking, Surrey RH5 6NT, UK\\
$^{6}$Dipartimento di Fisica e Astronomia Galileo Galilei, Universit\'a di Padova, via F. Marzolo 8, I-35131 Padova, Italy\\ 
$^{7}$Scuola Universitaria Superiore IUSS Pavia, Palazzo del Broletto, piazza della Vittoria 15, I-27100 Pavia, Italy\\
$^{8}$INAF--Istituto di Astrofisica Spaziale e Fisica Cosmica di Milano, via A.\,Corti 12, I-20133 Milano, Italy\\
$^{9}$Istituto Nazionale di Fisica Nucleare (INFN), Sezione di Pavia, via A.\,Bassi 6, I-27100 Pavia, Italy \\
$^{10}$Department of Physics, Università di Cagliari, S.P. Monserrato-Sestu km 0,700, I-09042 Monserrato, Italy \\
}

\date{Accepted XXX. Received YYY; in original form ZZZ}

\pubyear{2022}

\begin{document}
\label{firstpage}
\pagerange{\pageref{firstpage}--\pageref{lastpage}}
\maketitle

\begin{abstract}

The magnetar \srclong\ underwent a new active episode on 2020 April 27--28, when a forest of hundreds of X-ray bursts and a large enhancement of the persistent flux were detected. For the first time, a radio burst with properties similar to those of fast radio bursts and with a X-ray counterpart was observed from this source, showing that magnetars can power at least a group of fast radio bursts. In this paper, we report on the X-ray spectral and timing properties of \srclong\ based on a long-term monitoring campaign with \cxo, \xmm, \nustar, \swift\ and \nicer\ covering a time span of $\sim7$ months since the outburst onset. The broadband spectrum exhibited a non-thermal power-law component ($\Gamma \sim 1.2$) extending up to $\sim20-25$\,keV throughout the campaign and a blackbody component with temperature decreasing from $\sim1.5$\,keV at the outburst peak to $\sim0.45$\,keV in the following months. We found that the luminosity decay is well described by the sum of two exponential functions, reflecting the fast decay ($\sim$1\,d) at the early stage of the outburst followed by a slower decrease ($\sim$30\,d). The source reached quiescence about $\sim80$\,days after the outburst onset, releasing an energy of $\sim 6 \times 10^{40}$\,erg during the outburst. We detected X-ray pulsations in the \xmm\ data sets and derived an average spin-down rate of $\sim3.5 \times 10^{-11}$\,\ss\ using the spin period measurements derived in this work and three values reported previously during the same active period. Moreover, we report on simultaneous radio observations performed with the Sardinia Radio Telescope. No evidence for periodic or single-pulse radio emission was found.
\end{abstract}




\section{Introduction}

Among isolated neutron stars, magnetars are the most active, with a distinctive high-energy phenomenology \citep[see, e.g.,][for recent reviews]{kaspi17,esposito21}. Powered by their own magnetic energy, which is stored in a superstrong field (up to $\sim$10$^{15}$\,G at the surface), these objects emit X-ray/gamma-ray bursts that last from milliseconds to tens of minutes and reach a wide range of X-ray peak luminosities, 10$^{39}$ -- 10$^{45}$\,\lum. These flaring events are often accompanied by long-lived (up to years) enhancements of the persistent X-ray luminosity, the so-called outbursts \citep[see \url{http://magnetars.ice.csic.es};][]{cotizelati18}.          

Discovered in 2014 \citep{stamatikos14}, \srclong\ (henceforth \src) has a spin period $P\sim3.25$\,s and a spin-down rate $\dot{P} \sim 1.4\times10^{-11}$\,\ss, implying a surface dipolar magnetic field $B_{\rm p} \sim 2.2\times10^{14}$\,G at the pole \citep{israel16}. 
Since its discovery, \src\ has been one of the most active magnetars, showing outbursts in 2015 February, 2016 May and June, and frequent bursting episodes \citep[see, e.g.,][]{younes17,lin20}. Its latest reactivation dates back to 2020 April 27, when several X-ray and gamma-ray instruments detected a burst storm and an increase of the persistent X-ray flux \citep[e.g.,][]{palmer20, younes20}. A day after the initial trigger, the Canadian Hydrogen Intensity Mapping Experiment (CHIME) and the Survey for Transient Astronomical Radio Emission 2 (STARE2) independently detected an extremely bright radio burst \citep{chime20, bochenek20}, with morphology reminiscent of that of Fast Radio Bursts \citep[FRBs; see, e.g.,][for a review]{caleb21}. The energy released was about three orders of magnitude larger than that of any radio pulse from the Crab pulsar (the source emitting the brightest Galactic radio pulses; \citealt{bera19}) and any giant pulse detected from the radio magnetar XTE\,J1810--197 \citep{caleb22}, and $\sim$50 times smaller than that released by the weakest extragalactic FRB observed so far \citep[e.g.,][]{marcote20}. This detection strengthened the hypothesis that at least a sub-group of FRBs can be powered by magnetars at cosmological distances \citep{beloborodov17, margalit20}. Moreover, the radio burst was temporally coincident with a hard X-ray burst \citep{mereghetti20, tavani21, ridnaia21}, showing for the first time that magnetar bursts can have a bright radio counterpart. Furthermore, analysis of simultaneous radio and X-ray archival observations of magnetars revealed two FRB-like bursts from another source, 1E\,1547.0--5408 \citep{israel21}. One of the radio bursts was anticipated by $\sim$1\,s by a short X-ray burst, resulting in a radio-to-X-ray fluence ratio of $\sim$10$^{-9}$, proving that magnetars can emit radio bursts with fluences spanning over a wide range.  

No pulsed radio emission from \src\ was detected in the immediate aftermath of the FRB-like event \citep{lin20nature}. Coordinated radio and X-ray monitoring campaigns were initiated. While no other simultaneous radio and X-ray bursts were observed, \src\ emitted a few more fainter radio bursts \citep{kirsten20, zhang20} and several X-ray short bursts (see \url{http://enghxmt.ihep.ac.cn/bfy/331.jhtml} and Table\,2 by \citealt{borghese20}). On 2020 October 8, CHIME detected three additional radio bursts from the direction of \src, all clustered within one rotational period cycle \citep{good20}. Follow-up observations with the Five-hundred-metre Aperture Sperical Telescope (FAST) caught numerous single pulses from the source and also detected pulsed radio emission \citep{zhu20}. These detections indicated that \src\ can emit radio bursts with energies spanning nearly seven orders of magnitude and switch on/off in the radio band.

Here, we report on the results of the X-ray long-term monitoring campaign of \src\ covering the first $\sim$7 months of the outburst decay since its reactivation on 2020 April 27. We first summarise the data analysis procedure in Section\,\ref{sec:obs}. We then present the timing and spectral analysis, as well as a search for short bursts in Section\,\ref{sec:analysis}. Simultaneous radio observations are described in Section\,\ref{sec:radio}. Finally, we discuss our findings in Section\,\ref{sec:discuss}.

\begin{landscape}
\begin{table}
\begin{center}
\caption{Log of the X-ray and radio observations of \src\ analysed in this work.
\label{tab:new_obs}}

\begin{tabular}{@{}ccccccccc}
\hline 
\hline
X-ray Instrument$^a$ & Obs.ID & Start & Stop & Exposure &  Count Rate$^b$ & $kT_{\rm BB}$ & $R_{\rm BB}$ & Flux$^c$ \\
 & & \multicolumn{2}{c}{YYYY-MM-DD hh:mm:ss (TT)} & (ks) & (counts\,s$^{-1}$) & (keV) & (km) & (10$^{-12}$\,cgs) \\
\hline
\cxo/ACIS (TE) & 22431 & 2020-04-30 20:30:28 & 2020-05-01 02:44:51 & 19.8 & 0.147$\pm$0.003 & 0.51$\pm$0.03 & 2.1$\pm$0.2 & 4.6$\pm$0.4 \\
\cxo/ACIS (TE) & 22432 & 2020-05-02 08:58:14 & 2020-05-02 23:29:51 & 49.9 & 0.137$\pm$0.002 & 0.56$\pm$0.01 & 1.34$\pm$0.05 & 4.0$\pm$0.1 \\
\xmm/EPIC-pn (FF) & 0871190201 & 2020-05-13 21:43:24 & 2020-05-14 10:54:26 & 28.3 & 0.388$\pm$0.004 & 0.45$\pm$0.01 & 1.7$\pm$0.1 & 2.5$\pm$0.1 \\
\cxo/ACIS (TE) & 23251 & 2020-05-18 10:48:14 & 2020-05-18 16:32:19 & 18.8 & 0.116$\pm$0.002 & 0.50$\pm$0.01 & 1.5$\pm$0.1 & 3.0$\pm$0.1 \\

\nicer/XTI & 3655010201 & 2020-05-18 05:36:06 & 2020-05-18 13:38:40 & 4.7 & 0.29$\pm$0.01 & 0.49$\pm$0.02 & 1.3$\pm$0.2 & 0.9$\pm$0.1 \\
\nicer/XTI & 3020560105$^{d1}$ & 2020-05-19 21:53:47 & 2020-05-19 22:10:40 & 0.9 & 0.50$\pm$0.03 & 0.51$_{-0.03}^{+0.01}$ & 1.5$\pm$0.2 & 1.7$\pm$0.1 \\
\nicer/XTI & 3020560106$^{d1}$ & 2020-05-20 07:39:40 & 2020-05-20 15:37:20 & 0.6 & 0.67$\pm$0.04 & 0.51$_{-0.03}^{+0.01}$ & 1.5$\pm$0.2 & 1.7$\pm$0.1 \\
\nicer/XTI & 3020560107 & 2020-05-22 00:59:19 & 2020-05-22 22:54:46 & 5.1 & 0.49$\pm$0.01 & 0.47$\pm$0.02 & 1.5$\pm$0.1 & 2.1$\pm$0.1 \\
\nicer/XTI & 3020560108 & 2020-05-23 00:13:19 & 2020-05-23 09:42:50 & 3.2 & 0.39$\pm$0.02 & 0.45$\pm$0.02 & 1.8$\pm$0.2 & 1.0$\pm$0.1 \\
\nicer/XTI & 3020560109$^{d2}$ & 2020-05-25 14:13:00 & 2020-05-25 15:55:32 & 0.8 & 0.53$\pm$0.03 & 0.44$\pm$0.02 & 1.8$\pm$0.1 & 1.8$\pm$0.1 \\

\swift/XRT (PC) & 00033349067 & 2020-05-28 12:10:39 & 2020-05-28 16:59:54 & 2.0 & 0.024$\pm$0.004 & 0.37$_{-0.07}^{+0.09}$ & 1.8$_{-0.6}^{+2.0}$ & 1.9$_{-0.3}^{+0.4}$ \\

\nicer/XTI & 3020560110$^{d2}$ & 2020-05-28 21:10:39 & 2020-05-28 23:00:55 & 1.7 & 0.43$\pm$0.02 & 0.44$\pm$0.02 & 1.8$\pm$0.1 & 1.8$\pm$0.1 \\
\nicer/XTI & 3020560111$^{d3}$ & 2020-05-29 03:22:38 & 2020-05-29 03:39:46 & 0.9 & 0.38$\pm$0.03 & 0.48$\pm$0.04 & 1.5$\pm$0.02 & 1.7$\pm$0.1 \\
\nicer/XTI & 3020560112$^{d3}$ & 2020-05-30 05:42:03 & 2020-05-30 18:18:57 & 1.3 & 0.49$\pm$0.03 & 0.48$\pm$0.04 & 1.5$\pm$0.02 & 1.7$\pm$0.1 \\
\nicer/XTI & 3020560113$^{d4}$ & 2020-05-31 01:50:42 & 2020-05-31 11:24:40 & 1.1 & 0.49$\pm$0.03 & 0.43$\pm$0.02 & 1.6$\pm$0.02 & 2.6$\pm$0.2 \\
\nicer/XTI & 3020560114$^{d4}$ & 2020-06-01 02:37:42 & 2020-06-01 21:50:40 & 2.7 & 0.43$\pm$0.02 & 0.43$\pm$0.02 & 1.6$\pm$0.02 & 2.6$\pm$0.2 \\
\nicer/XTI & 3020560115$^{d4}$ & 2020-06-02 23:34:38 & 2020-06-02 23:51:40 & 1.0 & 0.47$\pm$0.03 & 0.43$\pm$0.02 & 1.6$\pm$0.02 & 2.6$\pm$0.2 \\ 
\nicer/XTI & 3020560116$^{d5}$ & 2020-06-03 04:12:40 & 2020-06-03 04:25:13 & 0.7 & 0.41$\pm$0.04 & 0.50$\pm$0.02 & 1.2$\pm$0.1 & 0.9$\pm$0.1 \\
\nicer/XTI & 3020560117$^{d5}$ & 2020-06-04 12:45:40 & 2020-06-04 13:10:23 & 1.4 & 0.37$\pm$0.02 & 0.50$\pm$0.02 & 1.2$\pm$0.1 & 0.9$\pm$0.1 \\

\swift/XRT (PC) & 00033349068 & 2020-06-05 03:37:29 & 2020-06-05 09:53:53 & 1.5 & 0.021$\pm$0.004 & 0.4$_{-0.1}^{+0.2}$ & 1.6$_{-0.7}^{+5.7}$ & 2.6$_{-0.7}^{+0.8}$  \\

\nicer/XTI & 3020560118$^{d5}$ & 2020-06-05 04:14:56 & 2020-06-05 16:59:23 & 1.2 & 0.30$\pm$0.02 & 0.50$\pm$0.02 & 1.2$\pm$0.1 & 0.9$\pm$0.1 \\
\nicer/XTI & 3020560119$^{d5}$ & 2020-06-06 05:03:57 & 2020-06-06T05:22:00 & 1.0 & 0.20$\pm$0.02 & 0.50$\pm$0.02 & 1.2$\pm$0.1 & 0.9$\pm$0.1 \\ 

\swift/XRT (PC) & 00033349069 & 2020-06-11 01:14:49 & 2020-06-11 14:16:52 & 2.5 & 0.023$\pm$0.003 & 0.55$_{-0.09}^{+0.10}$ & 0.9$_{-0.2}^{+0.4}$ & 2.9$\pm$0.4 \\

\nustar\ FPMA/B & 80602313006$^e$ & 2020-06-14 06:46:09 & 2020-06-14 23:21:09 & 30.1/29.4 & 0.073$\pm$0.002 & 0.44$\pm$0.04 & $2.0_{-0.5}^{+1.0}$ & 2.0$\pm$0.1  \\
\swift/XRT (PC) & 00089040001$^e$ & 2020-06-14 10:25:15 & 2020-06-1413:49:53 & 1.9 & 0.029$\pm$0.004 & 0.44$\pm$0.04 & $2.0_{-0.5}^{+1.0}$  & 2.0$\pm$0.1  \\

\nicer/XTI & 3655010301$^{d6}$ & 2020-06-17 21:38:39 & 2020-06-17 22:06:40 & 1.2 & 0.33$\pm$0.03 & 0.46$\pm$0.01 & 1.4$\pm$0.1 & 1.4$\pm$0.1  \\   

\swift/XRT (PC) & 00033349070 & 2020-06-18 08:23:18 & 2020-06-18 23:11:53 & 2.7 & 0.029$\pm$0.003 & 0.50$_{-0.05}^{+0.06}$ & 1.4$_{-0.3}^{+0.4}$  & 1.4$\pm$0.2 \\

\nicer/XTI & 3655010302$^{d6}$ & 2020-06-18 08:29:43 & 2020-06-18 21:21:20 & 6.6 & 0.38$\pm$0.01 & 0.46$\pm$0.01 & 1.4$\pm$0.1 & 1.4$\pm$0.1 \\
\nicer/XTI & 3655010303$^{d6}$ & 2020-06-18 23:59:32 & 2020-06-19 09:45:00 & 4.1 & 0.21$\pm$0.01 & 0.46$\pm$0.01 & 1.4$\pm$0.1 & 1.4$\pm$0.1 \\ 
\nicer/XTI & 3020560120$^{d7}$ & 2020-06-20 21:05:40 & 2020-06-20 21:23:40 & 0.9 & 0.46$\pm$0.03 & 0.42$\pm$0.01 & 1.8$\pm$0.2 & 1.7$\pm$0.1  \\
\nicer/XTI & 3020560121$^{d7}$ & 2020-06-21 00:01:20 & 2020-06-21 00:29:20 & 0.8 & 0.31$\pm$0.03 & 0.42$\pm$0.01 & 1.8$\pm$0.2 & 1.7$\pm$0.1  \\
\nicer/XTI & 3020560122$^{d7}$ & 2020-06-22 10:17:00 & 2020-06-22 10:49:20 & 1.7 & 0.41$\pm$0.02 & 0.42$\pm$0.01 & 1.8$\pm$0.2 & 1.7$\pm$0.1  \\
\nicer/XTI & 3020560123 & 2020-06-23 01:36:00 & 2020-06-23 17:48:20 & 3.2 & 0.31$\pm$0.01 & 0.45$\pm$0.01 & 1.5$\pm$0.1 & 0.8$\pm$0.1 \\
\nicer/XTI & 3020560124$^{d8}$ & 2020-06-24 22:41:19 & 2020-06-24 23:13:11 & 1.8 & 0.35$\pm$0.02 & 0.47$\pm$0.02 & 1.5$\pm$0.01 & 1.0$\pm$0.1  \\  

\swift/XRT (PC) & 00033349071 & 2020-06-25 01:30:52 & 2020-06-25 17:43:52 & 3.2 & 0.025$\pm$0.003 & 0.38$\pm$0.06 & 1.9$_{-0.5}^{+1.2}$  & 1.6$_{-0.3}^{+0.4}$  \\

\nicer/XTI & 3020560125$^{d8}$ & 2020-06-25 21:57:00 & 2020-06-25 22:29:00 & 1.6 & 0.37$\pm$0.03 & 0.47$\pm$0.02 & 1.5$\pm$0.01 & 1.0$\pm$0.1 \\

\hline
\end{tabular}
\begin{list}{}{}
\item[$^a$]The instrumental setup is indicated in brackets: TE = timed exposure, FF = full frame, PC = photon counting, WT = windowed timing.
\item[$^b$]Count rate, computed after removing bursts, in the 0.3--10\,keV range for \swift\ and \xmm, in the 0.3--8\,keV interval for \cxo, in the 1--5\,keV band for \nicer, and in the 3--20\,keV range for \nustar\ summing up the two FPMs. Uncertainties are at 1$\sigma$ c.l.
\item[$^c$]Observed 0.3--10\,keV flux in units of 10$^{-12}$\,\flux.
\item[$^d$,$^e$,$^f$] The spectra extracted from these observations were fitted jointly, tying up all model parameters (see Section~\ref{sec:phase_ave} for details).
\end{list}
\end{center}
\end{table}
\end{landscape}

\begin{landscape}
\begin{table}
\begin{center}
\contcaption{}
\label{tab:new_obs}
\begin{tabular}{@{}ccccccccc}
\hline 
\hline
X-ray Instrument$^a$ & Obs.ID & Start & Stop & Exposure &  Count Rate$^b$ & $kT_{\rm BB}$ & $R_{\rm BB}$ & Flux$^c$ \\
 & & \multicolumn{2}{c}{YYYY-MM-DD hh:mm:ss (TT)} & (ks) & (counts\,s$^{-1}$) & (keV) & (km) & (10$^{-12}$\,cgs) \\
\hline
\nicer/XTI & 3020560126$^{d9}$ & 2020-06-27 14:29:00 & 2020-06-27 14:45:20 & 0.8 & 0.38$\pm$0.03 & 0.45$\pm$0.03 & 1.5$\pm$0.2 & 1.1$\pm$0.1 \\
\nicer/XTI & 3020560127$^{d9}$ & 2020-06-28 16:31:00 & 2020-06-28 17:04:51 & 1.7 & 0.29$\pm$0.02 & 0.45$\pm$0.03 & 1.5$\pm$0.2 & 1.1$\pm$0.1 \\
\nicer/XTI & 3020560128$^{d9}$ & 2020-06-29 18:53:20 & 2020-06-29 19:24:40 & 1.7 & 0.31$\pm$0.02 & 0.45$\pm$0.03 & 1.5$\pm$0.2 & 1.1$\pm$0.1  \\
\nicer/XTI & 3020560129$^{d10}$ & 2020-06-30 16:19:20 & 2020-06-30 17:03:20 & 2.1 & 0.29$\pm$0.02 & 0.46$\pm$0.01 & 1.5$\pm$0.1 & 0.9$\pm$0.1 \\
\nicer/XTI & 3020560130$^{d10}$ & 2020-07-01 12:25:57 & 2020-07-01 12:40:50 & 0.6 & 0.41$\pm$0.04 & 0.46$\pm$0.01 & 1.5$\pm$0.1 & 0.9$\pm$0.1  \\
\nicer/XTI & 3020560131$^{d10}$ & 2020-07-02 11:41:29 & 2020-07-02 12:03:20 & 1.2 & 0.27$\pm$0.02 & 0.46$\pm$0.01 & 1.5$\pm$0.1 & 0.9$\pm$0.1 \\

\swift/XRT (PC) & 00033349072 & 2020-07-02 16:24:24 & 2020-07-02 18:27:52 & 3.2 & 0.026$\pm$0.003 & 0.51$_{-0.07}^{+0.08}$ & 1.2$_{-0.2}^{+0.5}$  & 3.4$\pm$0.4 \\

\nicer/XTI & 3020560132$^{d10}$ & 2020-07-03 14:01:00 & 2020-07-03 14:22:40 & 1.1 & 0.37$\pm$0.02 & 0.46$\pm$0.01 & 1.5$\pm$0.1 & 0.9$\pm$0.1 \\
\nicer/XTI & 3020560133$^{d10}$ & 2020-07-04 19:30:37 & 2020-07-04 19:47:06 & 0.8 & 0.33$\pm$0.03 & 0.46$\pm$0.01 & 1.5$\pm$0.1 & 0.9$\pm$0.1 \\

\nicer/XTI & 3020560134$^{d11}$ & 2020-07-08 18:21:40 & 2020-07-08 18:32:57 & 0.5 & 0.31$\pm$0.05 & 0.43$\pm$0.02 & 1.6$\pm$0.2 & 0.7$\pm$0.1 \\

\swift/XRT (PC) & 00033349073 & 2020-07-09 11:17:01 & 2020-07-09 21:01:52 & 3.4 & 0.027$\pm$0.003 & 0.44$\pm$0.05 & 1.6$_{-0.3}^{+0.5}$ & 1.4$\pm$0.2 \\

\nicer/XTI & 3020560135$^{d11}$ & 2020-07-10 05:43:00 & 2020-07-10 05:50:07 & 0.4 & 0.28$\pm$0.04 & 0.43$\pm$0.02 & 1.6$\pm$0.2 & 0.7$\pm$0.1 \\
\nicer/XTI & 3020560136$^{d11}$ & 2020-07-11 10:02:40 & 2020-07-11 11:47:40 & 1.3 & 0.26$\pm$0.03 & 0.43$\pm$0.02 & 1.6$\pm$0.2 & 0.7$\pm$0.1 \\
\nicer/XTI & 3020560137$^{d11}$ & 2020-07-12 20:07:20 & 2020-07-12 20:19:40 & 0.7 & 0.31$\pm$0.03 & 0.43$\pm$0.02 & 1.6$\pm$0.2 & 0.7$\pm$0.1 \\
\nicer/XTI & 3020560138$^{d12}$ & 2020-07-15 16:02:00 & 2020-07-15 19:35:40 & 3.1 & 0.23$\pm$0.02 & 0.42$\pm$0.01 & 1.8$\pm$0.1 & 0.7$\pm$0.1 \\
\nicer/XTI & 3020560139$^{d12}$ & 2020-07-16 06:07:56 & 2020-07-16 17:17:00 & 1.4 & 0.32$\pm$0.02 & 0.42$\pm$0.01 & 1.8$\pm$0.1 & 0.7$\pm$0.1 \\

\swift/XRT (PC) & 00033349074 & 2020-07-16 12:20:27 & 2020-07-16 23:36:52 & 1.5 & 0.026$\pm$0.004 & 0.55$_{-0.09}^{+0.10}$ & 1.0$_{-0.2}^{+0.5}$  & 1.2$\pm$0.2  \\

\nicer/XTI & 3020560140$^{d12}$ & 2020-07-17 03:22:40 & 2020-07-17 03:40:40 & 0.9 & 0.12$\pm$0.03 & 0.42$\pm$0.01 & 1.8$\pm$0.1 & 0.7$\pm$0.1 \\
\nicer/XTI & 3020560141$^{d12}$ & 2020-07-19 00:23:40 & 2020-07-19 00:40:20 & 0.8 & 0.33$\pm$0.03 & 0.42$\pm$0.01 & 1.8$\pm$0.1 & 0.7$\pm$0.1 \\

\swift/XRT (PC) & 00033349075 & 2020-07-21 22:49:04 & 2020-07-21 23:02:54 & 0.8 & 0.034$\pm$0.006 & 0.4$\pm$0.1 & 1.5$_{-0.6}^{+2.4}$ & 1.1$\pm$0.3  \\ 

\swift/XRT (WT) & 00033349076 & 2020-07-24 00:01:53 & 2020-07-24 01:46:56 & 2.9 & 0.02$\pm$0.01 & --  & --  & -- \\

\swift/XRT (PC) & 00033349077 & 2020-07-30 04:17:06 & 2020-07-30 20:35:54 & 2.4 & 0.024$\pm$0.003 & 0.53$_{-0.08}^{+0.07}$ & 1.1$_{-0.2}^{+0.5}$  & 2.4$\pm$0.6  \\

\nicer/XTI & 3020560143$^{d13}$ & 2020-07-31 19:03:58 & 2020-07-31 20:50:03 & 1.3 & 0.33$\pm$0.02 & 0.43$\pm$0.03 & 1.7$_{-0.1}^{+0.3}$ & 1.2$\pm$0.3 \\
\nicer/XTI & 3020560144$^{d13}$ & 2020-08-01 16:45:40 & 2020-08-01 17:12:36 & 1.4 & 0.32$\pm$0.02 & 0.43$\pm$0.03 & 1.7$_{-0.1}^{+0.3}$ & 1.2$\pm$0.3 \\
\nicer/XTI & 3020560146$^{d14}$ & 2020-08-03 09:22:40 & 2020-08-03 18:48:54 & 1.9 & 0.52$\pm$0.02 & 0.39$\pm$0.03 & 1.7$_{-0.2}^{+0.3}$ & 3.3$\pm$0.3  \\

\swift/XRT (PC) & 00033349078 & 2020-08-04 01:59:14 & 2020-08-04 02:07:53 & 0.5 & 0.013$\pm$0.005 & 0.7$_{-0.2}^{+0.5}$ & 0.5$_{-0.2}^{+0.6}$ & 6.9$_{-6.7}^{+31.4}$ \\

\nicer/XTI & 3020560147$^{d14}$ & 2020-08-04 20:40:00 & 2020-08-04 21:07:36 & 0.8 & 0.34$\pm$0.03 & 0.39$\pm$0.03 & 1.7$_{-0.2}^{+0.3}$ & 3.3$\pm$0.3 \\

\swift/XRT (PC) & 00033349079 & 2020-08-06 06:40:52 & 2020-08-06 23:07:52 & 2.9 & 0.015$\pm$0.002 & 0.26$_{-0.05}^{+0.08}$ & 4.8$_{-1.9}^{+8.7}$  & 0.8$_{-0.1}^{+0.2}$  \\

\nicer/XTI & 3020560148$^{d14}$ & 2020-08-06 12:56:23 & 2020-08-06 16:27:29 & 2.2 & 0.37$\pm$0.02 & 0.39$\pm$0.03 & 1.7$_{-0.2}^{+0.3}$ & 3.3$\pm$0.3 \\

\swift/XRT (PC) & 00033349080 & 2020-08-13 18:54:37 & 2020-08-13 19:07:52 & 0.8 & 0.020$\pm$0.005 & 0.5$\pm$0.2 & 1.2$_{-0.4}^{+2.3}$ & 2.3$_{-1.1}^{+1.2}$  \\
\swift/XRT (PC) & 00033349081 & 2020-08-15 20:17:15 & 2020-08-15 23:51:54 & 2.4 & 0.027$\pm$0.003 & 0.44$\pm$0.07 & 1.6$_{-0.4}^{+1.0}$ & 2.2$\pm$0.6  \\
\swift/XRT (PC) & 00033349082 & 2020-08-25 17:37:39 & 2020-08-25 19:35:53 & 2.9 & 0.021$\pm$0.003 & 0.40$_{-0.06}^{+0.07}$ & 1.7$_{-0.4}^{+1.0}$ & 1.3$_{-0.2}^{+0.3}$ \\

\nicer/XTI & 3020560149$^{d15}$ & 2020-08-28 21:22:25 & 2020-08-28 21:41:20 & 0.3 & 0.54$\pm$0.06 & 0.44$\pm$0.04 & 1.6$\pm$0.2 & 1.4$\pm$0.5  \\

\swift/XRT (PC) & 00033349083 & 2020-09-05 00:49:21 & 2020-09-05 05:33:54 & 2.0 & 0.027$\pm$0.003 & 0.43$_{-0.04}^{+0.05}$ & 1.7$_{-0.3}^{+0.6}$  & 1.4$\pm$0.2  \\

\swift/XRT (WT) & 00033349084 & 2020-09-10 22:39:20 & 2020-09-10 23:59:56 & 1.5 & 0.05$\pm$0.01 & --  & --  & -- \\ 
\nicer/XTI & 3020560151$^{d15}$ & 2020-09-11 05:55:00 & 2020-09-11 07:51:00 & 2.1 & 0.32$\pm$0.02 & 0.44$\pm$0.04 & 1.6$\pm$0.2 & 1.4$\pm$0.5  \\ 
\swift/XRT (WT) & 00033349085 & 2020-09-11 09:43:06 & 2020-09-11 11:15:56 & 2.2 & 0.034$\pm$0.008 & --  & --  & -- \\
\swift/XRT (WT) & 00033349086 & 2020-09-17 16:52:17 & 2020-09-17 23:20:56 & 3.3 & 0.049$\pm$0.006 & --  & --  & -- \\
\swift/XRT (WT) & 00033349087 & 2020-09-18 00:50:43 & 2020-09-18 02:34:56 & 1.7 & 0.039$\pm$0.008 & --  & --  & -- \\
\hline
\end{tabular}
\begin{list}{}{}
\item[$^a$]The instrumental setup is indicated in brackets: TE = timed exposure, FF = full frame, PC = photon counting, WT = windowed timing.
\item[$^b$]Count rate, computed after removing bursts, in the 0.3--10\,keV range for \swift\ and \xmm, in the 0.3--8\,keV interval for \cxo, in the 1--5\,keV band for \nicer, and in the 3--20\,keV range for \nustar\ summing up the two FPMs. Uncertainties are at 1$\sigma$ c.l.
\item[$^c$]Observed 0.3--10\,keV flux in units of 10$^{-12}$\,\flux.
\item[$^d$,$^e$,$^f$] The spectra extracted from these observations were fitted jointly, tying up all model parameters (see Section~\ref{sec:phase_ave} for details).
\end{list}
\end{center}
\end{table}
\end{landscape}

\begin{landscape}
\begin{table}
\begin{center}
  \contcaption{}
\begin{tabular*}{\textwidth}{c @{\extracolsep{\fill}}ccccccccc}
\hline 
\hline
X-ray Instrument$^a$ & Obs.ID & Start & Stop & Exposure &  Count Rate$^b$ & $kT_{\rm BB}$ & $R_{\rm BB}$ & Flux$^c$ \\
 & & \multicolumn{2}{c}{YYYY-MM-DD hh:mm:ss (TT)} & (ks) & (counts\,s$^{-1}$) & (keV) & (km) & (10$^{-12}$\,cgs) \\
\hline 
\nicer/XTI & 3020560152$^{d16}$ & 2020-09-19 01:50:25 & 2020-09-19 02:13:40 & 1.2 & 0.30$\pm$0.02 & 0.50$\pm$0.04 & 1.2$\pm$0.2 & 1.3$\pm$0.4 \\
\swift/XRT (WT) & 00033349088 & 2020-09-19 21:18:08 & 2020-09-19 23:07:56 & 1.5 & 0.06$\pm$0.01 & --  & --  & -- \\
\nicer/XTI & 3020560153$^{d16}$ & 2020-09-25 05:05:56 & 2020-09-25 09:55:17 & 1.4 & 0.36$\pm$0.02 & 0.50$\pm$0.04 & 1.2$\pm$0.2 & 1.3$\pm$0.4 \\
\xmm/EPIC-pn (FF) & 0871191301$^f$ & 2020-10-01 17:22:30 & 2020-10-02 16:05:13 & 55.7 & 0.244$\pm$0.002 & 0.44$\pm$0.01 & 1.6$\pm$0.1 & 1.3$\pm$0.1  \\
\nustar\ FPMA/B & 80602313008$^f$ & 2020-10-04 06:31:09 & 2020-10-05 04:51:09 & 40.6/40.2 & 0.063$\pm$0.002 & 0.44$\pm$0.01 & 1.6$\pm$0.1 & 1.3$\pm$0.1 \\ 
\nicer/XTI & 3655010401$^{d17}$ & 2020-10-06 01:24:13 & 2020-10-06 23:32:20 & 9.1 & 0.46$\pm$0.01 & 0.45$\pm$0.01 & 1.4$\pm$0.1 & 1.7$\pm$0.1 \\
\nicer/XTI & 3655010402$^{d17}$ & 2020-10-07 00:17:19 & 2020-10-07 11:27:40 & 14.6 & 0.30$\pm$0.01 & 0.45$\pm$0.01 & 1.4$\pm$0.1 & 1.7$\pm$0.1 \\

\swift/XRT (PC) & 00033349089 & 2020-10-08 21:26:39 & 2020-10-08 23:05:37 & 2.0 & 0.026$\pm$0.003 & 0.32$_{-0.06}^{+0.07}$ & 2.6$_{-0.9}^{+3.0}$  & 1.2$_{-0.3}^{+0.4}$ \\
\nicer/XTI & 3020560154$^{d18}$ & 2020-10-09 12:55:44 & 2020-10-09 22:19:00 & 0.9 & 0.37$\pm$0.03 & 0.46$\pm$0.02 & 1.5$\pm$0.1 & 1.0$\pm$0.1\\
\swift/XRT (WT) & 00033349090 & 2020-10-09 22:31:00 & 2020-10-10 00:19:00 & 2.5 & 0.32$\pm$0.01 & --  & --  & -- \\
\nicer/XTI & 3020560155$^{d18}$ & 2020-10-10 02:50:21 & 2020-10-10 21:33:20 & 1.9 & 0.37$\pm$0.02 & 0.46$\pm$0.02 & 1.5$\pm$0.1 & 1.0$\pm$0.1 \\
\swift/XRT (WT) & 00033349092 & 2020-10-11 16:07:14 & 2020-10-11 18:12:11 & 3.4 & 0.045$\pm$0.004 & --  & --  & -- \\

\swift/XRT (PC) & 00033349093 & 2020-10-12 03:20:14 & 2020-10-12 13:07:52 & 1.8 & 0.026$\pm$0.004 & 0.54$_{-0.08}^{+0.09}$ & 1.0$_{-0.2}^{+0.5}$  & 1.3$\pm$0.2      \\

\nicer/XTI & 3020560157$^{d18}$ & 2020-10-13 06:47:59 & 2020-10-13 19:24:51 & 2.1 & 0.31$\pm$0.02 & 0.46$\pm$0.02 & 1.5$\pm$0.1 & 1.0$\pm$0.1 \\

\swift/XRT (PC) & 00033349094 & 2020-10-13 13:01:52 & 2020-10-13 20:59:52 & 1 & 0.023$\pm$0.005 & 0.41$_{-0.07}^{+0.09}$ & 1.7$_{-0.5}^{+1.5}$  & 0.9$\pm$0.2  \\

\swift/XRT (PC) & 00033349095 & 2020-10-14 07:46:27 & 2020-10-14 22:24:52 & 0.9 & 0.025$\pm$0.005 & 0.4$\pm$0.1 & 1.4$\pm$0.5 & 1.4$_{-0.4}^{+0.8}$  \\

\nicer/XTI & 3020560159$^{d19}$ & 2020-10-16 21:40:00 & 2020-10-16 23:33:51 & 1.4 & 0.41$\pm$0.02 & 0.47$\pm$0.01 & 1.4$\pm$0.1 & 0.9$\pm$0.1 \\
\nicer/XTI & 3020560160$^{d19}$ & 2020-10-17 00:37:21 & 2020-10-17 08:36:42 & 11.1 & 0.34$\pm$0.01 & 0.47$\pm$0.01 & 1.4$\pm$0.1 & 0.9$\pm$0.1 \\
\nicer/XTI & 3020560161$^{d19}$ & 2020-10-18 18:13:20 & 2020-10-18 23:33:00 & 5.8 & 0.27$\pm$0.01 & 0.47$\pm$0.01 & 1.4$\pm$0.1 & 0.9$\pm$0.1 \\
\nicer/XTI & 3020560162$^{d20}$ & 2020-10-19 00:25:24 & 2020-10-19 22:46:00 & 13.5 & 0.37$\pm$0.01 & 0.47$\pm$0.01 & 1.4$\pm$0.1 & 1.2$\pm$0.1 \\
\nicer/XTI & 3020560163$^{d20}$ & 2020-10-19 23:40:02 & 2020-10-20 21:59:40 & 16.8 & 0.33$\pm$0.01 & 0.47$\pm$0.01 & 1.4$\pm$0.1 & 1.2$\pm$0.1 \\
\nicer/XTI & 3020560164$^{d21}$ & 2020-10-21 00:27:02 & 2020-10-21 22:46:00 & 6.6 & 0.35$\pm$0.01 & 0.48$\pm$0.01 & 1.3$\pm$0.1 & 1.3$\pm$0.1 \\
\nicer/XTI & 3020560165$^{d21}$ & 2020-10-21 23:44:35 & 2020-10-22 20:26:40 & 4.4 & 0.38$\pm$0.01 & 0.48$\pm$0.01 & 1.3$\pm$0.1 & 1.3$\pm$0.1 \\
\nicer/XTI & 3020560166$^{d22}$ & 2020-10-24 13:42:00 & 2020-10-24 23:22:12 & 7.4 & 0.48$\pm$0.01 & 0.44$\pm$0.02 & 1.3$\pm$0.1 & 3.5$\pm$0.2 \\
\nicer/XTI & 3020560167$^{d22}$ & 2020-10-25 11:20:59 & 2020-10-25 22:36:29 & 6.0 & 0.50$\pm$0.01 & 0.44$\pm$0.02 & 1.3$\pm$0.1 & 3.5$\pm$0.2 \\
\nicer/XTI & 3020560168$^{d23}$ & 2020-10-25 23:44:58 & 2020-10-26 23:25:40 & 8.4 & 0.46$\pm$0.01 & 0.45$\pm$0.02 & 1.2$\pm$0.1 & 3.3$\pm$0.1 \\
\nicer/XTI & 3020560169$^{d23}$ & 2020-10-28 12:12:01 & 2020-10-28 21:45:40 & 2.3 & 0.49$\pm$0.02 & 0.45$\pm$0.02 & 1.2$\pm$0.1 & 3.3$\pm$0.1 \\
\nicer/XTI & 3020560170$^{d24}$ & 2020-11-12 15:06:40 & 2020-11-12 21:36:12 & 2.6 & 0.47$\pm$0.01 & 0.48$\pm$0.01 & 1.3$\pm$0.1 & 0.9$\pm$0.1 \\
\nicer/XTI & 3020560171$^{d24}$ & 2020-11-13 12:39:00 & 2020-11-13 22:38:00 & 6.7 & 0.33$\pm$0.01 & 0.48$\pm$0.01 & 1.3$\pm$0.1 & 0.9$\pm$0.1 \\
\nicer/XTI & 3020560172$^{d25}$ & 2020-11-19 11:10:56 & 2020-11-19 22:26:49 & 4.4 & 0.32$\pm$0.01 & 0.44$\pm$0.01 & 1.5$\pm$0.1 & 1.8$\pm$0.1 \\
\nicer/XTI & 3020560173$^{d25}$ & 2020-11-20 11:55:00 & 2020-11-20 23:33:00 & 11.6 & 0.40$\pm$0.01 & 0.44$\pm$0.01 & 1.5$\pm$0.1 & 1.8$\pm$0.1 \\
\hline
\hline
\end{tabular*}
\begin{tabular*}{\textwidth}{c @{\extracolsep{\fill}}cccccccc}
Radio Instrument & Frequency & Bandwidth & Start & Stop & Exposure & Flux Density Upper Limit$^g$ & Fluence Upper Limit$^g$     \\
 & (GHz) & (MHz) & \multicolumn{2}{c}{YYYY-MM-DD hh:mm:ss (TT)} & (hr) & Periodic Emission (mJy) & Single Pulse (mJy\,ms)    \\
\hline
SRT & 1.5 & 460 & 2020-10-09 15:51:30 & 2020-10-09 19:04:12 & $2 \times 1.3$ & 0.1 & 800 & \\
SRT & 1.5 & 460 & 2020-10-10 16:31:12 & 2020-10-10 13:49:30 & $2 \times 1.3$ & 0.1 & 800 & \\
\hline
\end{tabular*}
\begin{list}{}{}
\item[$^a$]The instrumental setup is indicated in brackets: TE = timed exposure, FF = full frame, PC = photon counting, WT = windowed timing.
\item[$^b$]Count rate, computed after removing bursts,  in the 0.3--10\,keV range for \swift\ and \xmm, in the 0.3--8\,keV interval for \cxo, in the 1--5\,keV band for \nicer, and in the 3--20\,keV range for \nustar\ summing up the two FPMs.  Uncertainties are at 1$\sigma$ c.l.
\item[$^c$]Observed 0.3--10\,keV flux in units of 10$^{-12}$\,\flux.
\item[$^*$] These observations were merged in the spectral analysis.
\item[$^d$,$^e$,$^f$] The spectra extracted from these observations were fitted jointly, tying up all model parameters (see Section~\ref{sec:phase_ave} for details).
\item[$^g$] Upper limits are computed using the radiometer equation \citep{lorimer04}, assuming a pulse duty cycle of 5\%.
\end{list}
\end{center}
\end{table}
\end{landscape}

\section{X-ray observations and data reduction} 
\label{sec:obs}
Table\,\ref{tab:new_obs} reports a log of the X-ray observations of \src\ analysed in this work. These comprise three \cxo\ pointings (two of which unpublished) and one \xmm\ pointing carried out in 2020 between April 30 and May 18 (see also \citealt{gogus20}), and subsequent multi-instrument observations performed until 2020 November 20. These data sets complement those already presented by \citet{borghese20}, and provide a total time coverage spanning about 7 months since the source reactivation on 2020 April 27.  

Data reduction was performed using tools in the {\sc HEASoft} package (v.\,6.29c), the Science Analysis Software (v.\,19) and the \cxo\ Interactive Analysis of Observations (v.\,4.12) with the most recent calibration files. We referred photon arrival times to the Solar system barycenter using the \cxo\ position. (R.A. = 19$^\mathrm{h}$34$^\mathrm{m}$55$\fs$598, decl. = +21$^{\circ}$53$^{\prime}$47$\farcs$79, J2000.0; \citealt{israel16}) and the JPL planetary ephemeris DE\,200. In the following, we derive all quantities assuming a distance of 6.6\,kpc \citep{zhou20} and all uncertainties are quoted at 1$\sigma$ confidence level (c.l.). 

Diffuse emission, due to a scattering halo around the source, was detected at the outburst onset \citep{mereghetti20}. A detailed analysis of this component is beyond the scope of this work and will be presented in a future paper (Tiengo et al., in preparation). To avoid any contamination from the diffuse emission, we selected the background region far from the source (at an angular separation of at least 150 arcsec). 

\subsection{\swift} 

The \swift\ X-ray Telescope \citep[XRT;][]{burrows05} observed the source 29 times in 2020 between May 28 and October 14, with single exposures ranging from $\sim$0.5 to $\sim$3.4\,ks. The \swift/XRT was configured in photon counting (PC) mode in 21 observations, giving a readout time of about 2.5\,s. The remaining observations were performed in windowed timing mode (WT; readout time of 1.8\,ms). We reprocessed the data using standard prescriptions\footnote{See \url{https://www.swift.ac.uk/analysis/xrt/index.php}}. For the PC-mode observations, we extracted source photons from a circle centered on the source with a radius of 20 pixels, and background photons from an annulus with radii of 40 and 80 pixels, free of sources (1 XRT pixel corresponds to about 2\farcs36). For the WT-mode observations, we collected the source photons from a box of size 20$\times$40 pixels centered on the source, and estimated the background from a region of the same size located far from the source. Net count rates are listed in Table\,\ref{tab:new_obs}. Only the data sets collected in the PC-mode were used for the spectral analysis, whereas the WT-mode data were inspected only for the presence of short bursts.

\subsection{\xmm}

\src\ was observed with the European Photon Imaging Camera (EPIC) on board the \xmm\ satellite on 2020 May 13--14 and October 1--2 for an exposure time of 47.5\,ks and 81.5\,ks, respectively (for completeness, we included the observation ID 0871190201, already published by \citealt{gogus20}). In both pointings, the EPIC-pn \citep{struder01} was operating in Full Frame mode (FF; timing resolution of 73.4\,ms). The MOS cameras \citep{turner01} were set in FF mode (timing resolution of 2.6\,s) in the first observation and in Small Window mode (SW; timing resolution of 0.3\,s) during the second one. Here, we used only the data acquired with the EPIC-pn, which provides the data set with the highest counting statistics owing to its larger effective area compared to the MOS cameras. 

Standard analysis procedures were applied in the extraction of the scientific products. We cleaned the observations for periods of high background activity, resulting in a net exposure of 28.3\,ks and 55.7\,ks for the two observations. We collected the source photons from a circle of radius 30 arcsec. The background level was estimated from a 60-arcsec-radius circle far from the source, on the same CCD. We checked for the potential impact of pile-up through the \textsc{epatplot} tool and found a negligible pile-up fraction of $\sim$0.3\%. The response matrices and ancillary files were generated by means of the {\sc rmfgen} and {\sc arfgen} tasks, respectively. Background-subtracted and exposure-corrected light curves were extracted using {\sc epiclccorr}.

\subsection{\cxo}

Three observations of \src\ were carried out by \cxo\ using the Advanced CCD Imaging Spectrometer (ACIS; \citealt{garmire03}) since the onset of the latest outburst in 2020 April for a total on-source exposure time of 88.5\,ks. The ACIS was set in timed exposure (TE) mode with a frame readout time of 3.14\,s and the source was always positioned on the back-illuminated S3 chip. The timing resolution is too coarse to study the magnetar timing properties, therefore we include these data sets only in the spectral analysis. The observation ID 23251 was already presented by \citet{gogus20}. However, we re-analysed it consistently with our approach. 

Source photons were selected from a 1.5-arcsec circular region centered on the source, while the background counts were extracted from a circle with a radius of 40 arcsec far from the source. We estimated the impact of pile-up with \textsc{WebPimms}\footnote{\url{https://heasarc.gsfc.nasa.gov/cgi-bin/Tools/w3pimms/w3pimms.pl}.} and found that its fraction ranges from 13\% to 18\% across the different observations. Hence, pile-up is not negligible in our data. 
We created the source and background spectra, the associated redistribution matrices and ancillary response files using the {\sc specextract} script, and accounted for the effects of pile-up as explained in Section~\ref{sec:phase_ave}.

\subsection{\nustar} 

\nustar\ \citep{harrison13} observed \src\ at two epochs, in 2020 June and in 2020 October. The total on-source exposure times were 70.7\,ks and 69.6\,ks for the focal plane module A and B (FPMA and FPMB hereafter), respectively. We processed the event lists and filtered out passages of the satellite through the South Atlantic Anomaly using the tool {\sc nupipeline}. Both source and background counts were accumulated within a circular region of radius 100 arcsec. We then applied the script {\sc nuproducts} to extract light curves and spectra, and generate response files for both FPMs.
The source was detected up to $\sim$20\,keV in both observations at a net count rate of $\sim$0.07 counts\,s$^{-1}$ and $\sim$0.06 counts\,s$^{-1}$ in 2020 June and October (summing up the two FPMs), respectively. 

\subsection{\nicer}

The X-ray Timing Instrument (XTI) of the \nicer\ mission \citep{gendreau12} monitored \src\ intensively in 2020, starting from its reactivation on April 27. In this work, we focus on 71 observations, performed between 2020 May and November, for a total on-source exposure time of $\sim$220\,ks. The pointings acquired between 2020 April 28 and July 26 were already presented by \citet{younes20}. However, we decided to re-analyse them adopting a consistent approach with ours. 
 
We processed the data via the {\sc nicerdas} pipeline, using the {\sc nicerl2} tool and adopting standard filtering criteria. We created the ancillary response and response matrix files with the tools {\sc nicerarf} and {\sc nicerrmf}, respectively. The background count rates and spectra are computed through the {\sc nibackgen3C50} tool.

\section{X-ray analysis and results}\label{sec:analysis}

\subsection{Timing Analysis}\label{sec:timing} 

\begin{figure}
    \centering
    \includegraphics[width=1.\columnwidth, trim= 2cm 3.5cm 2cm 4cm, clip=true]{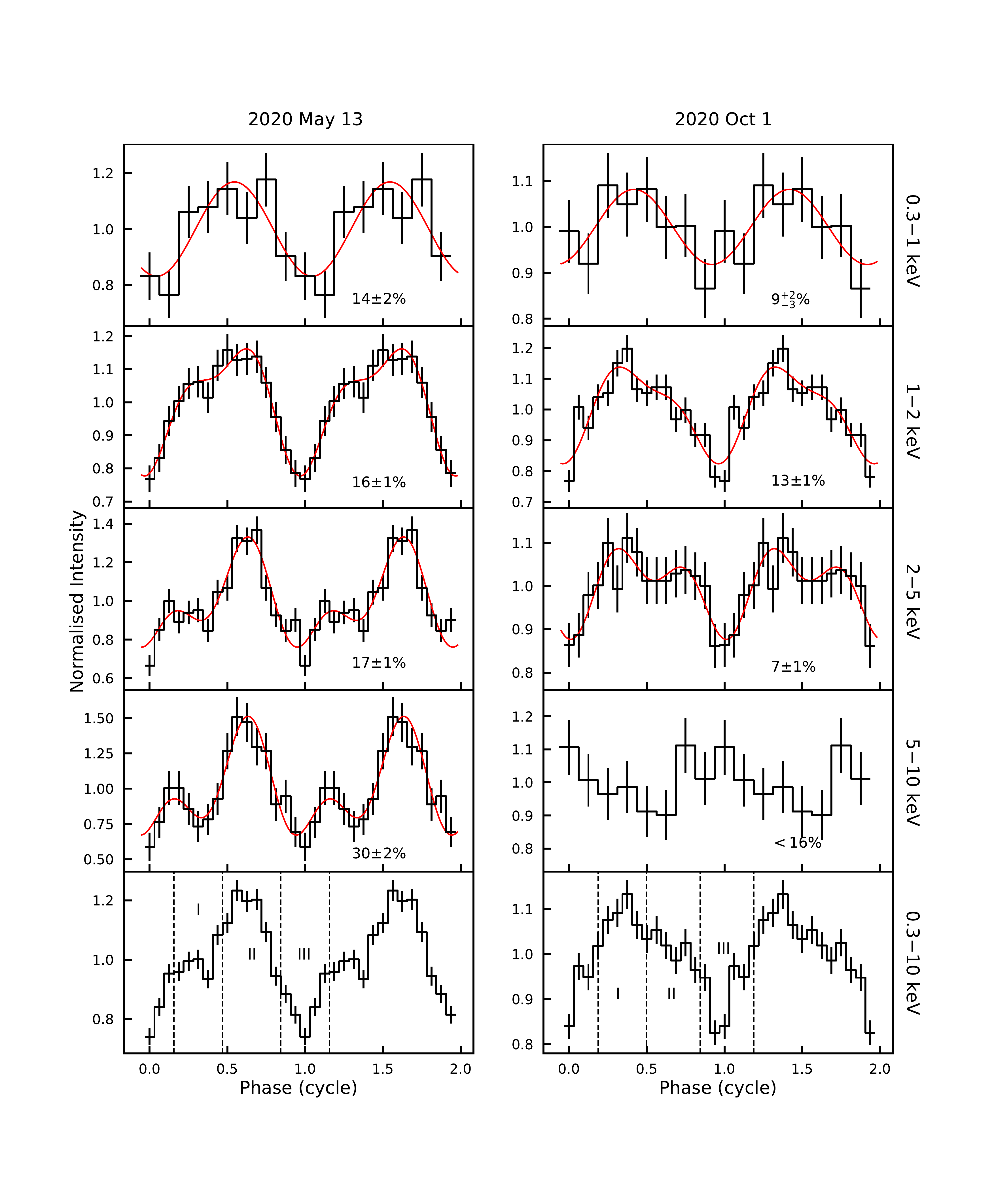}
    \vspace{-0.5cm}
    \caption{\xmm\ EPIC-pn background-subtracted, energy-resolved pulse profiles for the 2020 May (left-hand panel) and October (right-hand panel) data sets. In the bottom panel, the dashed lines indicate the intervals used in the phase-resolved spectral analysis. Two cycles are shown for clarity. }
    \label{fig:foldxmm}
 \end{figure}

The arrival times of the \xmm/EPIC-pn photons extracted from the source and background regions were corrected to the barycenter of the Solar system. Coherent pulsations were detected in the power spectra of both data sets at a high significance level ($>$11$\sigma$; \citealt{israel96}).
By means of a phase-fitting timing analysis in each pointing, we inferred the following best period: $P=3.247361(6)$\,s for observation ID 0871190201 (EPIC-pn data only; 2020 May 13) and $P=3.247778(6)$\,s for observation ID 0871191301 (EPIC-pn plus EPIC-MOSs merged data; 2020 October 1). The former period is in agreement,  within the uncertainties, with that reported by \cite{gogus20}.  
Figure\,\ref{fig:foldxmm} shows the corresponding light curves folded on the above periods as a function of energy for the two epochs. The pulse profile exhibits a quasi-sinusoidal shape below 1\,keV, well-fit by a sine function, and evolves to a more complex morphology with increasing energy, requiring a second sine to properly model the shape and displaying two peaks separated by $\sim$0.47 in phase in the 5--10\,keV range. The pulsed fraction (defined as the semi-amplitude of the fundamental divided by the average count rate) increased from (14$\pm$2)\% in the 0.3--1\,keV interval to (30$\pm2$)\% in the 5--10\,keV interval in 2020 May, while we detected pulsations till 5\,keV during the second observation. We set a 3$\sigma$ upper limit of $\sim$16\% in the 5--10\,keV band in 2020 October. Moreover, the broad-band (0.3--10\,keV) pulsed fraction dropped from (18$\pm$1)\% to (10$\pm$1)\% between the two epochs.

A similar procedure was followed for \nustar\ data sets (we selected photons in the 3--12\,keV and 3--5\,keV ranges) and the pulsar signal was searched in a narrow period interval around the value inferred from the \xmm\ data. No significant signal was detected and 3$\sigma$ upper limits in the 24\%--40\% range were obtained for the pulsed fraction. 

Based on the above inferred periods and those already reported by \cite{borghese20}, hence covering a time span of $\sim$5 months from 2020 April 28 till October 1, we inferred a first period derivative $\dot{P}$ = 3.5(1)$\times$10$^{-11}$\,s\,s$^{-1}$. This estimate is a factor 2.5 higher than the period derivative derived during the first four months of the 2014 outburst with a phase-coherent analysis \citep[$\dot{P}\sim1.4\times10^{-11}$\,s\,s$^{-1}$;][]{israel16}.

\subsection{Spectral Analysis}\label{sec:spec}

The spectral fitting was performed using {\sc Xspec} \citep{arnaud96}. We adopted the {\sc Tbabs} model with cross-sections of \citet{verner96} and elemental abundances of \citet{wilms00} to calculate the effects of the photoelectric absorption along the line of sight. Following our previous work \citep{borghese20}, we fixed the hydrogen column density to \nh\ = 2.3 $\times$ 10$^{22}$\,\cmdue, that is, the value inferred from a systematic analysis of high quality data acquired during the previous outbursts of \src\ (\citealt{cotizelati18}; see also \citealt{younes17}). 

Due to the low photon counting statistics, the \swift/XRT background-subtracted spectra were grouped to have at least five counts in each spectral channel and the W-statistic was employed for model parameter estimation and error calculation. For \nustar, \xmm, \cxo\ and \nicer, we binned the spectra to guarantee at least 50 background-subtracted counts per energy bin so as to use the \chisq\ statistics, unless otherwise specified.

\subsubsection{Phase-averaged spectral analysis}
\label{sec:phase_ave}

We fit the \xmm\ and \swift/XRT spectra in the 0.5--10\,keV energy range, and the \cxo\ and \nicer\ ones in the 0.3--8\,keV and 1--5\,keV intervals, respectively. For the \nustar\ pointings, the spectral analysis was limited to the 3--20\,keV energy band owing to the very low source signal-to-noise ratio above 20\,keV.

\begin{figure}
    \centering
    \includegraphics[width=1.\columnwidth]{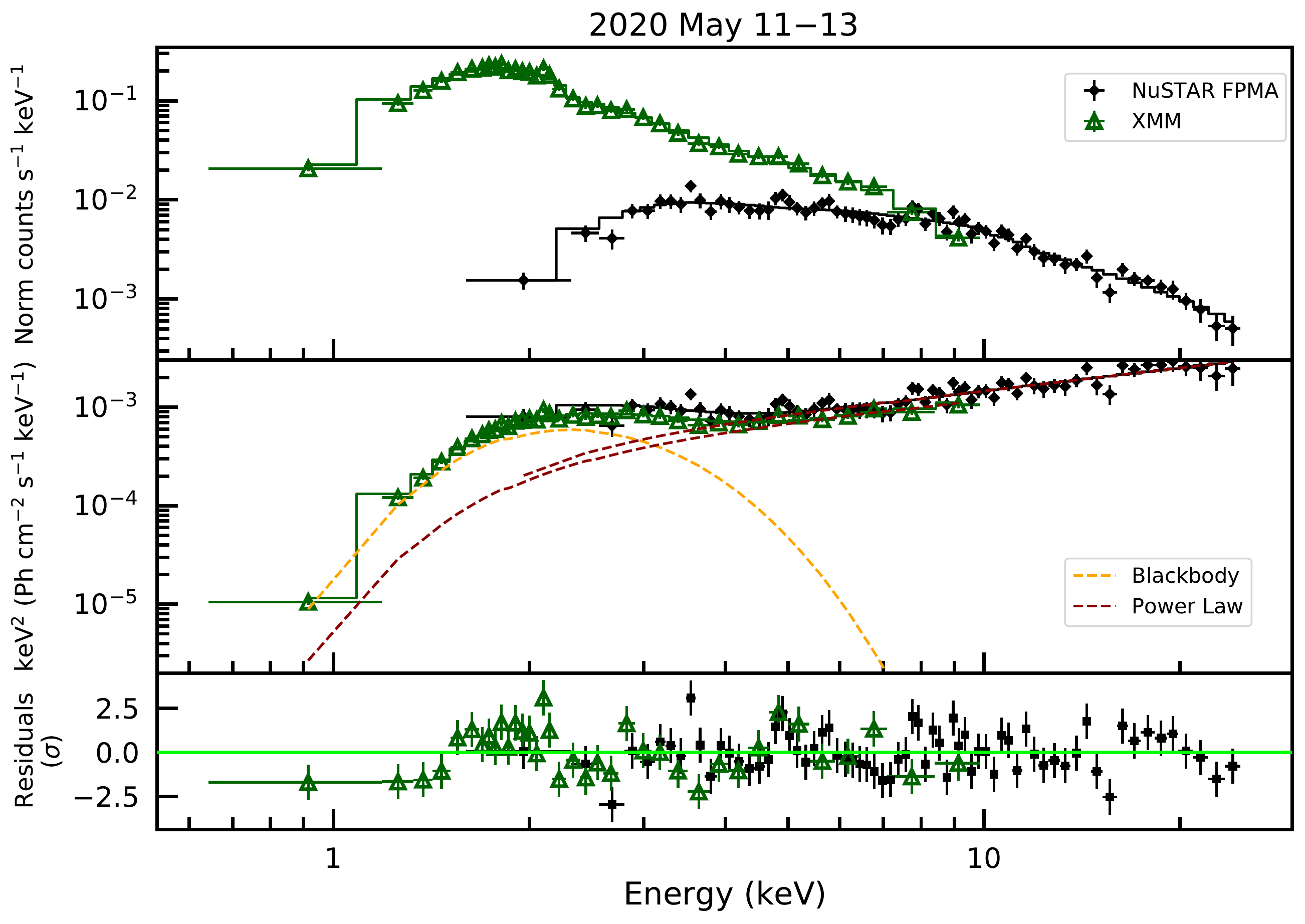}
    \includegraphics[width=1.\columnwidth]{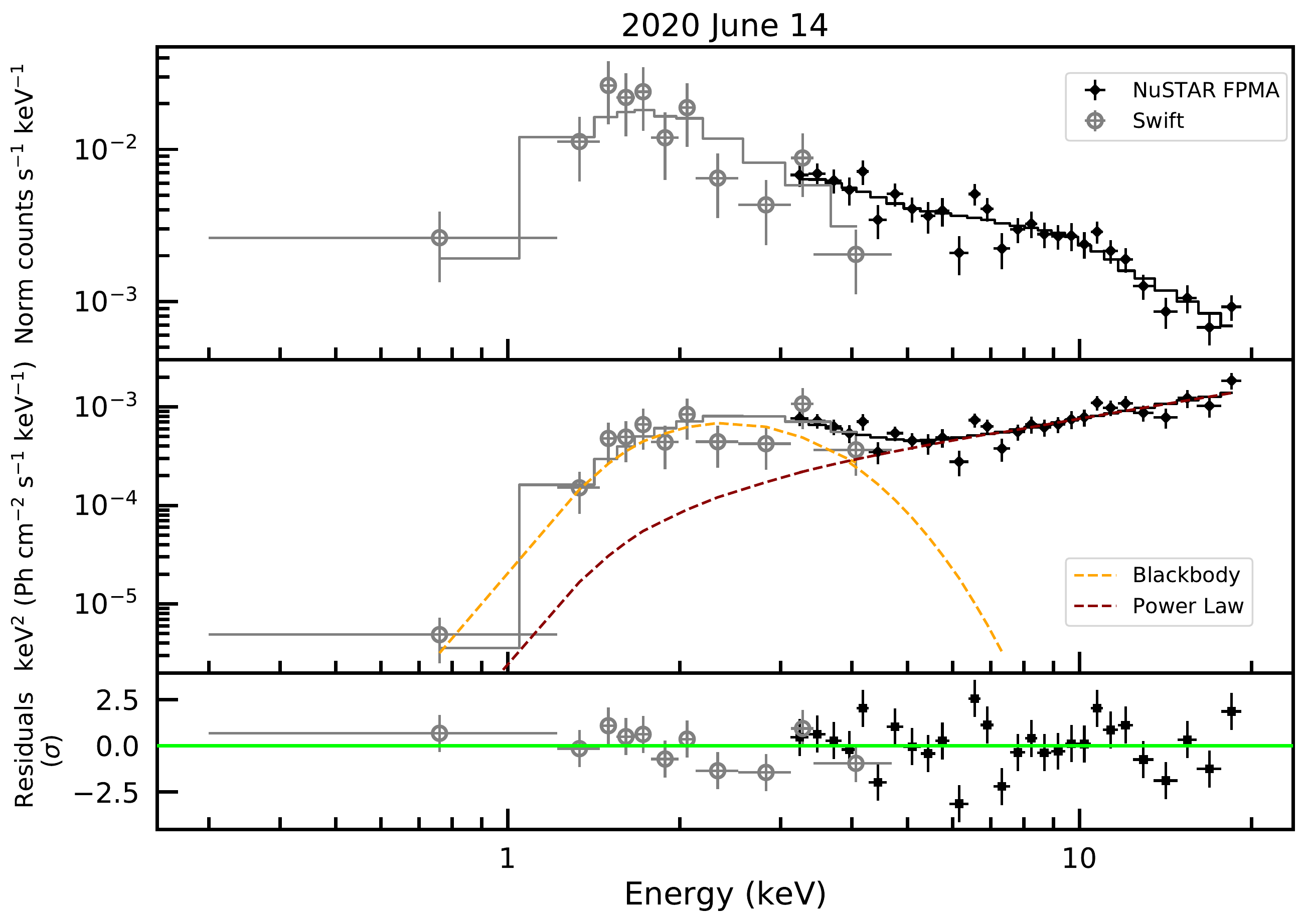}
    \includegraphics[width=1.\columnwidth]{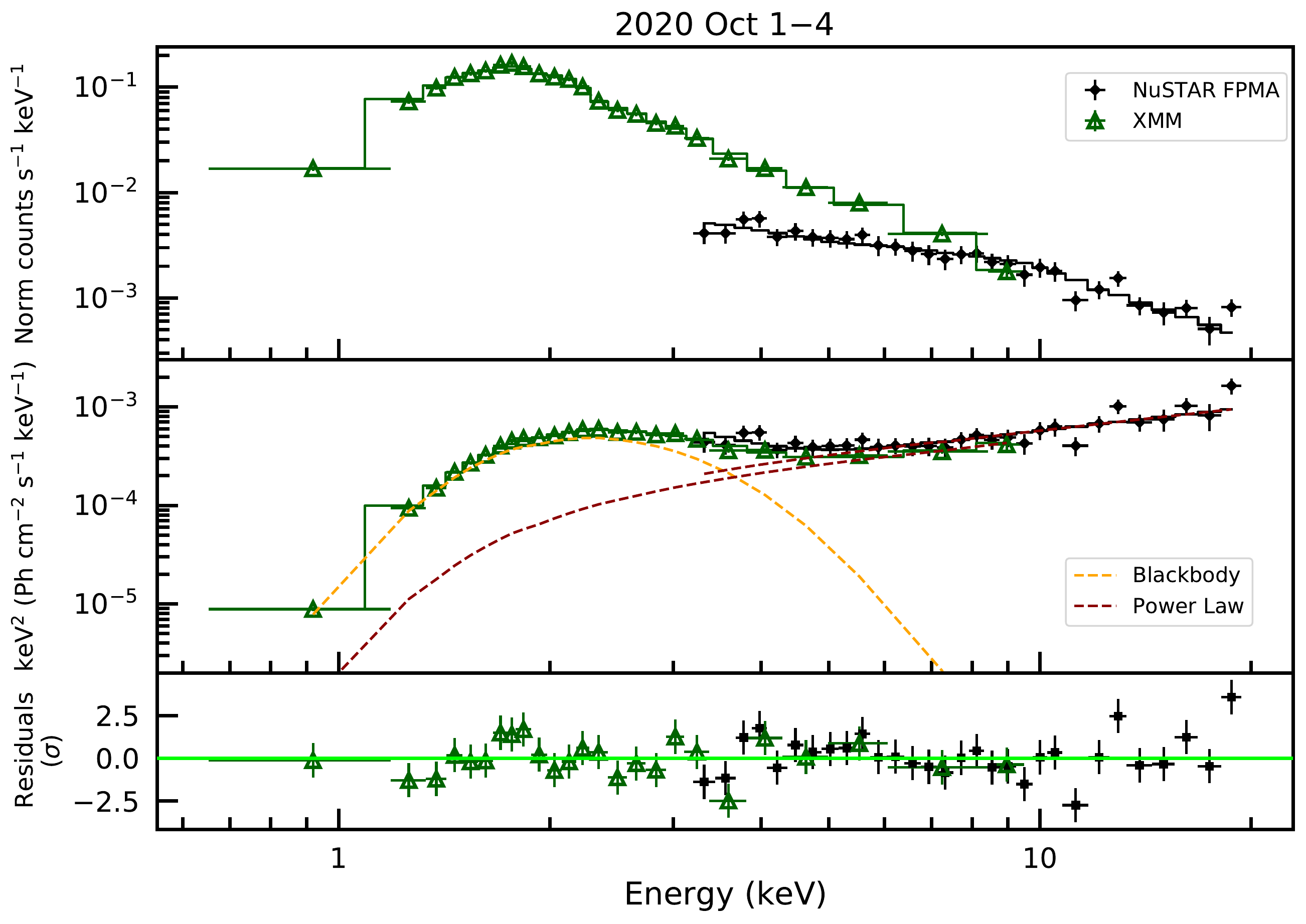}
    \caption{Broadband spectra of \src\ extracted from the quasi-simultaneous \swift/XRT (grey), \nustar\ (black) and \xmm\ (green) data on 2020 May 11--13 ({\it top}), June 14 ({\it middle}), and October 1--4 ({\it bottom}). \xmm\ data were re-binned for plotting purpose. For each plot, the $E^2 f(E)$ unfolded spectrum is shown in the middle panel. Dashed lines mark the contribution of the single components to the spectral model. Post-fit residuals in units of standard deviations are shown in the bottom panel. }
\label{fig:pha_ave_spec}
 \end{figure}

We start the fitting procedure by modelling the broadband spectra for the epochs 2020 May 11--13\footnote{We fit jointly the spectra extracted from the observations \xmm\ ID 0871190201 and \nustar\ ID 80602313004. The latter was already presented in our previous work \citep{borghese20}.}, June 14 and October 1--4. For each epoch, we fit the spectra jointly, forcing the model parameters to be the same across the data sets. Moreover, we include a multiplicative normalization which was frozen to one for the \nustar/FPMA spectrum, and allowed to vary for the other instruments. This term takes into account cross-calibration uncertainties between different instruments. Similarly as in the early stages of the outburst \citep{borghese20}, the spectra are well described by an absorbed blackbody plus a power-law component (BB+PL; see Figure\,\ref{fig:pha_ave_spec}). The best-fitting parameters are listed in Table\,\ref{tab:spec_ave}, where we also include the results of the broadband spectral analysis of the observations performed close to the outburst onset (2020 May 2). The 0.3--20\,keV luminosity decreased from (4.0 $\pm$ 0.3) $\times$ 10$^{34}$ $d^2_{6.6}$\,\lum\ on 2020 May 2 to (2.3 $\pm$ 0.1) $\times$ 10$^{34}$ $d^2_{6.6}$\,\lum\ on Oct 1--4, with a contribution of the power-law component of $\sim$75\% and $\sim$45\% respectively, in the same energy band ($d_{6.6}$ is the source distance in units of 6.6\,kpc). We did not detect a clear time evolution of the photon index $\Gamma$, which attained a value of $\sim$1.2, and of the radius of the thermal emitting region, with an average value of $R_{\rm BB} \sim$1.4\,km. On the other hand, the PL normalization decreased by a factor of $\sim$2.5 and $kT_{\rm BB}$ decreased from 0.59$_{-0.05}^{+0.06}$\,keV to 0.44$\pm$0.01\,keV during the time span covered by the broadband observations (2020 May 2 -- October 4). 

We then fit the same model to the \swift/XRT spectra jointly, freezing the \nh\ to the above value. We allowed the other parameters to vary, although the photon index was not constrained over the energy range covered by \swift. Hence, we fixed this parameter to $\Gamma$=1.2, that is, the averaged value derived from the broadband spectral analysis including \nustar\ spectra. 
The same procedure was applied to the \nicer\ spectra. To increase the source signal-to-noise, we fitted together \nicer\ spectra extracted from observations performed a few days apart, tying up all model parameters (see Table\,\ref{tab:new_obs}). Given the limited energy interval (1--5\,keV), $\Gamma$ was frozen to 1.2 in this fit as well. We obtained a W-stat = 856.55 for 881 degrees of freedom (d.o.f.) for the \swift\ data and \chisq/ = 3092.3 for 2495 d.o.f. for the \nicer\ data sets.

For the \cxo\ spectra, we estimated a pile-up fraction of 13--18\%. To correct for this effect, we included the multiplicative pile-up model \citep{davis01}, as implemented in \textsc{Xspec}, in the spectral fitting procedure. Following `The \cxo\ ABC guide to Pileup'\footnote{See \url{http://cxc.harvard.edu/ciao/download/doc/pileup_abc.pdf}.}, we allowed the grade migration parameter $\alpha$ to vary and fixed the parameter {\it psffrac} equal to 0.95, that is, we assumed that 95\% of events are contained within the central, piled-up portion of the source point spread function. We fit simultaneously the three spectra adopting an absorbed BB+PL model corrected by the pile-up model. As before, we fixed \nh=2.3 $\times$ 10$^{22}$\,\cmdue\ and $\Gamma$=1.2. The fit yielded \chisq\ = 177.83 for 181 d.o.f.

\begin{figure*}
    \centering
    \includegraphics[width=2.\columnwidth, trim=0 0 25cm 0, clip]{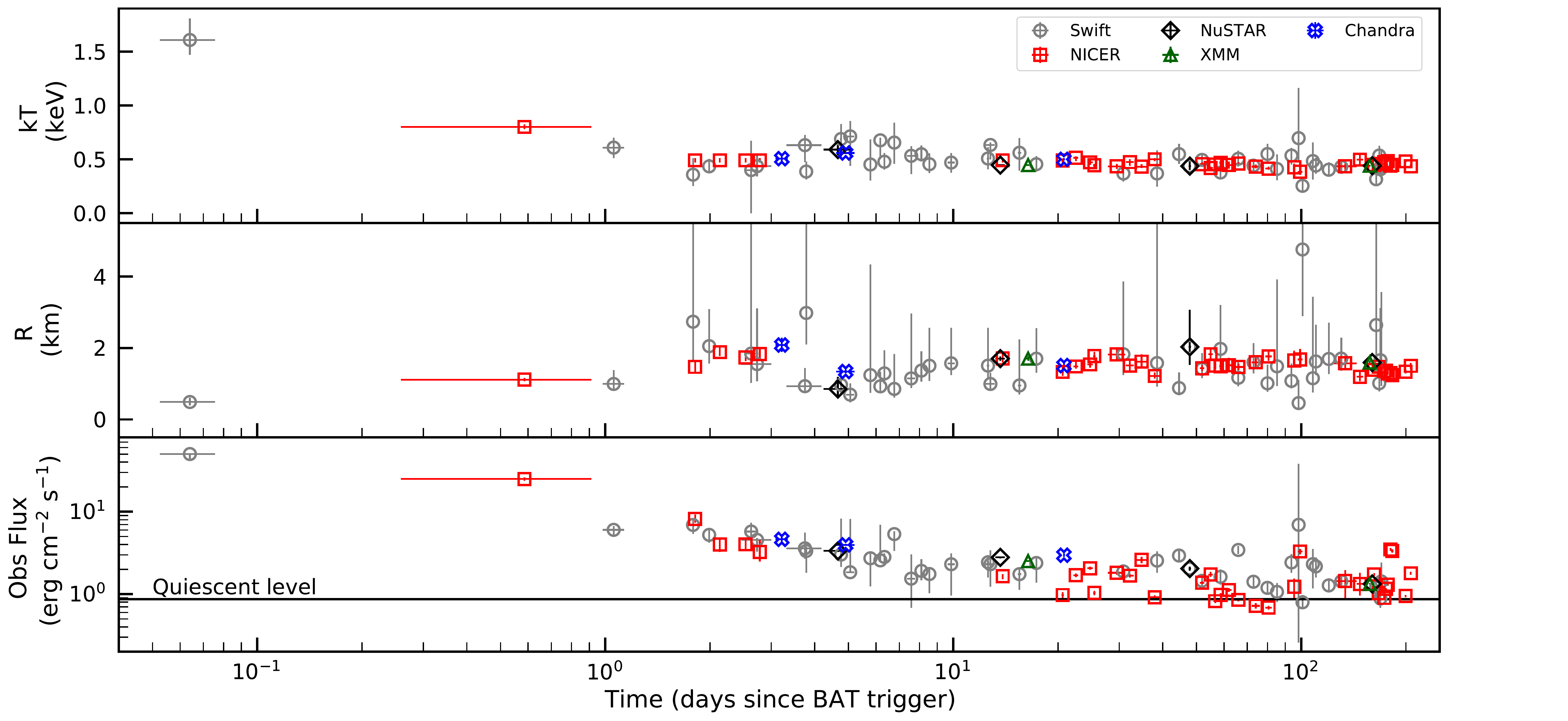}
    \caption{Time evolution of the blackbody temperature (top) and radius (middle) and of the observed flux (bottom) of \src\ between 2020 April 27 and November 20, including observations already presented by \citet{borghese20}. The observed fluxes are computed in the energy range 0.3--10\,keV and are expressed in units of 10$^{-12}$~\flux. The solid line in the bottom panel marks the quiescent flux, (8.7$\pm$0.3)$\times$10$^{-13}$\flux.}
    \label{fig:spec_evo}
 \end{figure*}

The best-fitting values for the radius and temperature of the blackbody component and the total observed flux (0.3--10\,keV) corresponding to each observation are reported in Table\,\ref{tab:new_obs}. Figure\,\ref{fig:spec_evo} shows the temporal evolution of these quantities. After an initial rapid decrease over the course of a few days, the blackbody temperature settled on a steady value of $\sim$0.45\,keV; while the blackbody radius did not show a strong variability, attaining an average value of $\sim$1.6\,km during the $\sim$200 days covered by the monitoring campaign. These values are consistent with those derived for the BB component during the previous outbursts, when the soft ($<$10\,keV) spectra were described by a BB+PL model and did not show any spectral variability, except for the brightness \citep{younes17}. Moreover, they are compatible also with the averaged spectral parameters of the BB component ($kT_{\rm BB} \sim$ 0.45\,keV and $R_{\rm BB} \sim$ 1.5\,km) as obtained from \nicer\ observations carried out during 2017--2019 \citep{younes20}.

\subsubsection{Quiescent level}
\label{sec:quiescence}
The quiescent level of \src\ is still unknown. In our previous work \citep{borghese20}, we assumed the value derived from the \swift/XRT observation performed 4 days before the outburst onset (ID 00033349044; 2020 April 23). The source signal-to-noise ratio was not high enough to perform a sensitive spectral analysis, therefore we used {\sc WebPIMMS} to derive an estimate of the observed flux, $\sim$4.5$\times$10$^{-13}$\,\flux\ (0.3--10\,keV; assuming a BB spectrum with \nh=2.3$\times$10$^{22}$\,\cmdue\ and $kT_{\rm BB}$=0.5\,keV). \citet{younes20} used the average obtained from the \nicer\ 2017--2019 monitoring campaign as a flux reference value ($\sim$6.7$\times$10$^{-13}$\,\flux\ adopting a BB model with \nh=2.4$\times$10$^{22}$\,\cmdue, $kT_{\rm BB}$=0.45\,keV and unabsorbed flux of 2$\times$10$^{-12}$\,\flux; 0.3--10\,keV). 
Finally, \citet{cotizelati18} performed a systematic spectral study of the major magnetar outbursts occurred up to the end of 2016 and identified the quiescent level of \src\ with the flux measured in the \xmm\ observation performed on 2014 October 4, (8.6$\pm$0.2)$\times$10$^{-13}$\,\flux\ (0.3--10\,keV; 2BB). \xmm/EPIC provides the most accurate characterization of the spectrum of \src\ at the faint flux levels observed outside the outburst episodes. Hence, we deem that the last observation provides the most reliable approximation to the true quiescent level for this source. To be consistent with our analysis, we re-fit the \xmm/EPIC-pn spectrum with a BB+PL model freezing \nh=2.3$\times$10$^{22}$\,\cmdue. The fit gave an overall satisfactory description with $kT_{\rm BB} = 0.48\pm0.01$\,keV, $R_{\rm BB} = 1.1\pm0.1$\,km and $\Gamma = 1.9^{+0.5}_{-1.4}$ (\chisq\ = 25.8 for 27 d.o.f.). We estimated the quiescent level of the observed flux to be equal to (8.7$\pm$0.3)$\times$10$^{-13}$\,\flux, that corresponds to a quiescent luminosity of (1.3$\pm$0.1)$\times$10$^{34}$\,\lum\ (0.3--10\,keV).

\begin{table*}
\begin{center}
\caption{Results of the joint spectral fits presented in Section \ref{sec:spec}.}
\label{tab:spec_ave}
\begin{tabular}{|l|ccccccc}
\hline
\hline
Epoch &  $kT_{\rm BB}$ & $R_{\rm BB}$ & $\Gamma$ & Norm PL & Flux$^a$ (Obs / Unabs) & Flux$^a$ Unabs BB / PL & \chisq/d.o.f. \\
      &  (keV)  &   (km)  &    & (pho keV$^{-1}$ cm$^{-2}$ s$^{-1}$)    & \multicolumn{2}{c}{(10$^{-12}$\,\flux)}& \\
\hline
2020 May 2$^b$ & 0.59$^{+0.06}_{-0.05}$ & 0.85$^{+0.35}_{-0.18}$ & 1.17$\pm$0.06 & (2.5$\pm$0.4)$\times$10$^{-4}$ & 5.8$\pm$0.1 / 7.8$\pm$0.6 & 2.2$\pm$0.5 / 5.6$\pm$0.2 & 148.4/139 \\ 

2020 May 11--13  & 0.45$\pm$0.01 & 1.7$\pm$0.1 & 1.24$\pm$0.04 & (2.5$\pm$0.2)$\times$10$^{-4}$ & 5.2$\pm$0.1 / 7.9$\pm$0.2 & 2.8$\pm$0.1 / 5.1$\pm$0.1 & 232.9/232 \\ 

2020 Jun 14$^b$ & 0.44$\pm$0.04 & 2.0$^{+1.0}_{-0.5}$ & 1.07$\pm$0.11 & (9.1$\pm$2.1)$\times$10$^{-5}$ & 3.4$\pm$0.1 / 6.2$\pm$0.5 & 3.7$\pm$1.3 / 2.5$\pm$0.1 & 53.17/49 \\

2020 Oct 1--4 & 0.44$\pm$0.01 & 1.6$\pm$0.1 & 1.23$\pm$0.07 & (9.9$\pm$1.4)$\times$10$^{-5}$ & 2.5$\pm$0.1 / 4.4$\pm$0.2 & 2.4$\pm$0.1 / 1.9$\pm$0.1  & 184.6/152 \\ 

\hline
\hline
\end{tabular}
\begin{list}{}{}
\item[$^a$]The fluxes are estimated in the 0.3--20\,keV energy range.
\item[$^b$]For these two epochs, the \nustar\ spectrum is fitted together with a \swift\ spectrum taken almost simultaneously. For the latter, the W-stat=7 for 7 d.o.f. for May 2 and W-stat=6.8 for 10 d.o.f. for Jun 14.
\end{list}
\end{center}
\end{table*}

\subsubsection{Phase-resolved spectroscopy}  
\label{sec:phase_res}

We performed a phase-resolved spectral analysis using the two \xmm/EPIC-pn data sets, where we detected the source spin period signal with high significance. We extracted spectra from three phase intervals (see Figure\,\ref{fig:foldxmm}, bottom panels) and fitted them using an absorbed BB+PL model. Similarly to the phase-averaged spectral analysis, the column density was held fixed at \nh\ = 2.3$\times$10$^{22}$\,\cmdue\ for all the fits. 

Firstly, for each epoch, we allowed only the normalizations of each component to vary, while the blackbody temperature and photon index were frozen at their best-fitting values for the corresponding phase-averaged spectrum (see Table\,\ref{tab:spec_ave}). The fit yielded \chisq\ = 206.9 for 159 d.o.f. and \chisq\ = 187.9 for 169 d.o.f. for the data sets acquired on 2020 May 13 and October 1, respectively. We obtained statistically equivalent fits by allowing all parameters to vary among the phase-resolved spectra, with \chisq\ = 205.1 for 153 d.o.f. (2020 May) and 179.8 for 163 d.o.f. (2020 October). The best-fitting values are listed in Table\,\ref{tab:spec_pps}. The blackbody radius and temperature do not show significant variations with the spin phase, while some hints for a phase dependence of the power-law index may be present.

\begin{table}
\begin{center}
\caption{Results of the phase-resolved spectral analysis presented in Section \ref{sec:phase_res}.}
\label{tab:spec_pps}
\begin{tabular}{l|ccc}
\hline
\hline
 &  \multicolumn{3}{c|}{2020 May 13} \\
\hline
 Phase     & $kT_{\rm BB}$ & $R_{\rm BB}$ & $\Gamma$  \\
          &  (keV)  &   (km)  &    \\
\hline
I & 0.47$\pm$0.01 & 1.6$\pm$0.1 & 0.9$^{+0.2}_{-0.3}$ \\
II & 0.45$\pm$0.01 & 1.5$\pm$0.1 & 1.0$\pm$0.2 \\ 
III & 0.43$\pm$0.02 & 1.6$\pm$0.1 & 1.4$\pm$0.2 \\
\hline
\hline
 & \multicolumn{3}{c}{2020 Oct 1} \\
 \hline 
Phase     & $kT_{\rm BB}$ & $R_{\rm BB}$ & $\Gamma$\\
          &  (keV)  &   (km)  &    \\
\hline
I & 0.44$\pm$0.01 & 1.5$\pm$0.1 & 1.2$\pm$0.3 \\
II & 0.46$\pm$0.01 & 1.4$\pm$0.1 & 0.9$\pm$0.3 \\
III & 0.43$\pm$0.01 & 1.5$\pm$0.1 & 1.2$\pm$0.3 \\
\hline
\hline
\end{tabular}
\end{center}
\end{table}

\subsection{Burst Search and Spectral Modelling}\label{sec:burst}

A search for short X-ray bursts was performed on all available data sets, using the same method outlined by \citet{borghese20} (see also, e.g., \citealt{gkw04}). We extracted the time series with time resolutions of 2.5073\,s for the \swift/XRT PC-mode data, of 73.36\,ms for the \xmm/EPIC-pn data sets, and of 1/16, 1/32 and 1/64\,s in all other cases. We labelled as bursts the bins with a probability $<$10$^{-4}$($NN_{\rm trials}$)$^{-1}$, where $N$ is the total number of time bins in a given light curve and $N_{{\rm trials}}$ corresponds to the number of timing resolutions used in the search. In Table\,\ref{tab:new_bursts}, we list the epochs of the bursts referred to the Solar system barycenter and Figure\,\ref{fig:bursts} displays their light curves.  

On 2020 October 8--9, \src\ entered a new radio active phase with the detection of multiple radio bursts and, for the first time, of pulsed radio emission at the X-ray spin period \citep{good20,zhu20}. On 2020 October 9, \swift/XRT observed the source for 2.5\,ks and we detected 24 bursts clustered within about ten minutes, corresponding to a burst rate of $\sim$0.04\,burst\,s$^{-1}$. Unfortunately, no meaningful spectral analysis could be carried out on the bursts detected using \swift/XRT owing to severe pile-up and saturation effects at the measured count rates. 

On the other hand, meaningful spectra could be extracted for a couple of bursts detected by \nicer, that is, those dubbed 3655010201 \#2 and 3655010301 \#1 in Table\,\ref{tab:new_bursts} (these are actually the events with the largest number of counts among the \nicer\ sample; see also Table\,\ref{tab:nicer_bursts}). For these events, the background level was estimated from the persistent emission detected in a close-in-time 10-s chunk of the data sets and the spectra were grouped to contain a minimum of 3 counts per spectral bin, allowing us to use the $W$-statistic. The averaged spectra of these bursts were then fitted with an absorbed power-law model, fixing the column density to \nh\ = 2.3 $\times$ 10$^{22}$\,\cmdue. We obtained the following best-fitting parameters: $\Gamma = -1.2\pm1.1$ and averaged unabsorbed flux $F_{{\rm X,unabs}} = 3.3_{-1.6}^{+3.2} \times10^{-8}$\,\flux\ (0.3--10\,keV) for burst 3655010201 \#2 (W-stat = 4.90 for 9 d.o.f.); $\Gamma = 2.0\pm$0.3 and $F_{{\rm X,unabs}} = 3.2_{-0.5}^{+0.7} \times10^{-8}$\,\flux\ (0.3--10\,keV) for burst 3655010301 \#1 (W-stat = 46.83 for 42 d.o.f.).

No significant bursts were detected in the \xmm/EPIC-pn light curves.

\section{Simultaneous radio observations}
\label{sec:radio}
\src\ was observed with the Sardinia Radio Telescope (SRT; \citealt{bolli,prandoni}) at 1.5 GHz in 2020 on October 9 and 10, simultaneously with \nicer, for two consecutive 1.3-hr sessions each day (see Table\,\ref{tab:new_obs} for details). Data were recorded with the ATNF digital backend PDFB3\footnote{See \url{http://www.srt.inaf.it/media/uploads/astronomers/dfb.pdf}} in search mode over a bandwidth of 460\,MHz split into 1\,MHz channels. Total intensity data were 2-bit sampled every 100\,$\mu$s, except for the first half of the October 10 run, where full Stokes data were recorded every 256\,$\mu$s (the backend showed signs of overheating and the previous configuration was restored for the second part of the run).

The data were folded (with the software {\sc dspsr}; \citealt{vanstraten11}) using the ephemeris obtained from X-ray data and using a dispersion measure DM = 332.8\,pc\,cm$^{-3}$ \citep{chime20}, and blindly searched (with the package {\sc presto}\footnote{\url{https://github.com/scottransom/presto}};  \citealt{ransom11}) over a DM range from 300 to 360\,pc\,cm$^{-3}$. A search spanning $\pm 0.1$\,ms around the nominal period of the pulsar and the same DM range as the blind search was done on the folded data using {\sc pdmp} (from the software package {\sc psrchive}; \citealt{hotan04}). No persistent radio pulsations were found down to a flux density limit of $\sim 0.1$\,mJy.

A search for single pulses was performed on the data using the {\sc spandak} pipeline\footnote{\url{https://github.com/gajjarv/PulsarSearch}} \citep{gajjar18}
The pipeline uses {\sc rfifind} from the {\sc presto} package for high-level radio frequency interference (RFI) excision.
The search for bursts/single pulses is conducted through {\sc Heimdall} \citep{bbb+12} to quickly search across a DM range from 0 to 1000\,pc\,cm$^{-3}$.
The de-dispersed time-series were searched for pulses using a matched-filtering technique with a maximum window size of 400 ms. Each candidate found by {\sc Heimdall} at DMs within the range 300--400\,pc\,cm$^{-3}$ was scrutinized against all other candidates for each given observation to validate and identify only the genuine ones.
A single candidate not resembling RFI was found at a DM compatible with that of the previously observed bursts (see e.g. \citealt{2021NatAs...5..414K}): DM = 332.42\,pc\,cm$^{-3}$. The candidate had a signal-to-noise ratio $S / N < 7$. In order to verify the genuineness of this candidate, a 1-s segment of data around the candidate has been reprocessed with an ad-hoc program with more sensitive RFI excision procedures taking into account lower level RFI.
Firstly, a search for the most corrupted frequency channels in the DM zero data was carried out using the spectral kurtosis algorithm \citep{7833535} as provided by the software package {\sc your}\footnote{\url{https://github.com/thepetabyteproject/your/}} \citep{Aggarwal2020}. We used a spectral kurtosis thresholding of $5\sigma$. Subsequently, we applied baseline subtraction and the data were normalized for the average bandpass. A check for possible corrupted temporal bins due to the presence of impulsive RFI was then performed with inter-quantile range (IQR) mitigation, similarly to \cite{10.1117/12.2559937}. The reprocessed data were then de-dispersed to the derived DM and smoothed via a 2-dimensional Gaussian filter. After this cleaning procedure, the candidate did not display any FRB-like characteristics either in the dynamic spectrum, or in the DM vs time plot, and we concluded that it was originated by RFI corrupted data.
With no candidates found, we set an upper limit of 800\,mJy on the fluence of a ms-long burst happening during these observations.

Since the data are uncalibrated, all upper limits reported above have been estimated using the modified radiometer equation (see e.g. \citealt{lorimer04}) adopting an antenna gain of 0.55\,K/Jy, a system temperature of 35\,K, a sky temperature of 25\,K (accounting also for the supernova remnant hosting \src), a threshold $S / N$ of 7, and, for the periodic emission, a duty cycle of 5\%. 

\section{Discussion} \label{sec:discuss}

On 2020 April 27, \src\ entered its fifth recorded outburst phase, placing itself in the short list of magnetars showing recurrent outbursts and frequent bursting activity, including e.g., 1E\,1048.1$-$5937, SGR\,1627$-$41 and \wes\ \citep{an18,borghese19,archibald20}. This latest outburst stood out from the previous events experienced by \src\ because it was accompanied by a remarkable X-ray burst forest (with more than 200 bursts detected in $\sim$20 minutes; \citealt{younes20}), and the emission of an intense radio burst with properties resembling those of FRBs and a X-ray counterpart \citep[e.g.,][]{chime20, mereghetti20}. 

Here we presented the temporal evolution of the spectral and timing properties of the source as tracked by an intensive X-ray monitoring campaign over $\sim$200 days since the outburst onset, as well as simultaneous radio observations.

\subsection{Light curve modelling}
To characterise the post-outburst luminosity decay, we modelled the temporal evolution of the 0.3--10\,keV luminosity with a phenomenological model consisting of a constant and two exponential functions:
\begin{equation}
    L(t) = L_{\rm q}+ \sum_{i=1}^{2} A_i \exp(-(t-t_0)/\tau_i)~,
\end{equation}
where $L_{\rm q}$ is the quiescent level, $t_{\rm 0}$ is the epoch of the outburst onset and the $e$-folding time $\tau$ can be considered as an estimate of the decay timescale.     
We fixed $t_{\rm 0}$ to MJD 58966.7683 (2020 April 27, 18:26:20 UTC), that is, the epoch at which \swift/BAT triggered on the first burst emitted from \src\ during this latest active period \citep{barthelmy2020}. For the quiescent luminosity, we assumed the value derived fitting a BB+PL model to the spectrum extracted from the \xmm\ observation performed on 2014 October 4, (1.3$\pm$0.1)$\times$10$^{34}$\,\lum\, (see Sec.\,\ref{sec:quiescence}). A 10 per cent error was assigned to each luminosity. The best-fitting values for the $e$-folding times are $\tau_1$=0.62$\pm$0.09\,d and $\tau_2$=31.2$\pm$3.5\,d, highlighting an initial fast decay followed by a slower decrease. The source reached quiescence about $\sim$80\,days after the outburst onset, releasing an energy of $\sim$5.8$\times$10$^{40}$\,erg. Note that the quiescent level of \src\ is not known yet and the true value could be lower than that assumed in this work. Therefore, the released energy and decay timescales should only be considered as a rough estimate. \citet{younes20} modelled the flux evolution over a period of three months after the outburst onset. Similar to our results, they found two decay trends described by very different $e$-folding times. The initial rapid decay is characterized by $\tau_1$=0.65$\pm$0.08\,d, which is consistent with our findings. However, the long-term flux decay has an $e$-folding time of $\tau_2$=75$\pm$5\,d, which differs from our results. The discrepancy might be due to the different quiescent level we assumed (see Sec.\,\ref{sec:quiescence}) and/or to the fact that our monitoring campaign extends over a longer period.

\citet{younes17} derived the energy emitted for the previous four outbursts within 10 days since the onset (see Table\,5 of their paper). The energy released in the first 10 days of the 2020 outburst is equal to $\sim$2.6$\times$10$^{40}$\,erg, second only to the 2016 June event when the energy was estimated to be $\sim$3.6$\times$10$^{40}$\,erg. For the first two outbursts in 2014 and 2015, the emitted energy in the first 10 days was $\sim$1.2$\times$10$^{40}$\,erg, while for the 2016 May episode it was slightly higher, $\sim$2$\times$10$^{40}$\,erg.

Overall, the values for the decay timescale and the total energy released for the latest outburst of \src\ fall at the low end of the range of values measured for magnetar outbursts. As a matter of fact, the decay time scale of $\sim$30\,d is among the shortest $\tau$ measured so far. Yet, they are still compatible with the trend of the correlation measured previously, according to which the shorter the outburst, the less energetic. These results imply that the decay pattern of this outburst is not dissimilar from those observed in other magnetars \citep{cotizelati18}.

\begin{table}
\begin{center}
\caption{Results of the broadband spectral fitting adopting the NTZ model (see Sec.\,\ref{sec:spec_disc}).}
\label{tab:spec_ave_ntz}
\begin{tabular}{|l|cccc}
\hline
\hline
Epoch & $kT$ & $\beta_{\rm bulk}$ & $\Delta \phi$ & Norm \\
 & (keV) &  & (rad) & \\
\hline
May 2 & 0.67$\pm$0.05 & 0.61$\pm$0.04 & 0.49$\pm$0.01 & 0.026$\pm$0.002 \\
May 11 & 0.48$\pm$0.01 & 0.65$\pm$0.02 & 0.452$\pm$0.002 & 0.034$\pm$0.001 \\
Jun14 & 0.50$\pm$0.03 & 0.72$\pm$0.06 & 0.43$\pm$0.01 & 0.030$\pm$0.005 \\
Oct 1 & 0.48$\pm$0.01 & 0.71$\pm$0.04 & 0.423$\pm$0.002 & 0.024$\pm$0.001 \\
\hline
\hline
\end{tabular}
\end{center}
\end{table}

\subsection{Spectral evolution}
\label{sec:spec_disc}

About five days after its reactivation, \src\ was observed with \nustar\ and \swift, revealing a hardening of the spectrum with the appearance of a non-thermal component extending up to $\sim$25\,keV. In the following months, three additional broadband observations were performed and still detected hard X-ray emission till $\sim$20\,keV. At each epoch, the non-thermal component was well modeled by a PL with a photon index of $\Gamma \sim$1.2. Its contribution to the total 0.3--20\,keV luminosity decreased from $\sim$75\% at the outburst peak to $\sim$45\% after $\sim$5 months (Table\,\ref{tab:spec_ave}). During the whole monitoring campaign, besides the PL component, a blackbody was required to properly model the spectrum. Its temperature rapidly decayed during the first day of the outburst from $\sim$1.5\,keV to $\sim$0.6\,keV and decreased only slightly down to $\sim$0.45\,keV over the following months. The corresponding emitting area was rather steady in time, with a radius of $\sim$1.6\,km (Figure\,\ref{fig:spec_evo}).

The spectral hardening and the detection of a power law at hard X-rays are ubiquitous properties of magnetars in outburst. The decomposition of the spectral model as a blackbody plus a power-law component is generally interpreted in terms of thermal emission from the cooling neutron star surface that gets affected by physical mechanisms taking place in the magnetosphere, such as Resonant Cyclotron Scattering \citep[RCS; see e.g.,][]{nobili08a}. The thermal photons produced at the surface gain energy via repeated scatterings onto charged particles flowing along the magnetic field lines, leading to the formation of a tail at higher energies. During an outburst, magnetic stresses and instabilities induce crustal displacements that can implant a strong twist of the magnetic field lines. The detection of a hot spot suggests that the magnetic twist is localized to a restricted portion of the magnetosphere, most likely to a current-carrying bundle of field lines \citep{beloborodov09}. 

However, explaining the spectral evolution of \src\ along the outburst within the RCS scenario poses some challenges. In fact, while the blackbody temperature quickly drops as expected \citep{beloborodov09,pons12}, the blackbody radius undergoes little changes and, even more strikingly, the power-law index remains almost constant. Actually, as recent 3D simulations have shown, the heated region can indeed cool without much shrinking \citep{degrandis21} but
the power-law should become softer as the twist subsides \citep{beloborodov09}. To investigate this further, we fitted the spectra with the NTZ model \citep{nobili08a, nobili08b}, which accounts for resonant cyclotron up-scattering of the soft seed photons (see Table\,\ref{tab:spec_ave_ntz}). Taken at face value, the results of the NTZ spectral fits seem to indicate that, while the luminosity of the source is decaying, the decrease in the twist angle $\Delta \phi$ is accompanied by an increase in the velocity $\beta_{\rm bulk}$ of the magnetospheric charges. Since both these quantities control the efficiency of the scattering process and hence the steepness of the power-law tail, this may in turn result in a nearly constant power-law index. The decrease in flux of the non-thermal component may reflect the fact that a smaller fraction of the photons from the thermal emitting area is intercepted by the currents in the bundle.

\subsection{Timing properties and pulse profile simulations}

Regarding the timing properties, we detected the spin period signal in the two \xmm\ data sets (2020 May 13 and October 1). The pulse profile displays a variable morphology with energy. In the 5--10\,keV interval, the profile exhibits a double-peaked shape, as observed in the previous \nustar\ observations performed close to the outburst onset \citep{borghese20}; while it evolves to a nearly sinusoidal shape at lower energies. The broadband pulsed fraction decreased by a factor of $\sim$2 between the two epochs. These behaviours are at odds with that observed during the first months of the 2014 outburst, when the pulse profile attained a quasi-sinusoidal shape, with no variation in time and energy, and the broadband pulsed fraction was in the 17--21\,\% range. These differences may suggest that distinct regions on the neutron star surface are heated during each outburst. By combining the two \xmm\ spin period measurements with those presented by \citet{borghese20} ($P$ = 3.24731(1)\,s on 2020 April 29--30, 3.247331(3)\,s on May 2 and 3.24731(1)\,s on May 11), we inferred a long-term average spin-down rate equal to $\dot{P}$ = 3.5(1)$\times$10$^{-11}$\ss, that is a factor 2.5 larger than the $\dot{P}$ measured in 2014 ($\dot{P}$=1.43(1)$\times$10$^{-11}$\ss; \citealt{israel16}). Changes in the pulse profile morphology and in the timing parameters are common during magnetar outbursts, mirroring the magnetosphere variations that follow flaring activity (for a more detailed discussion about the timing behavior of \src\ during 2020 October 1 and November 27 see Younes et al., to be subm.).    

\begin{figure}
    \centering
    \vspace{-6.cm}
    \includegraphics[width=1.1\columnwidth]{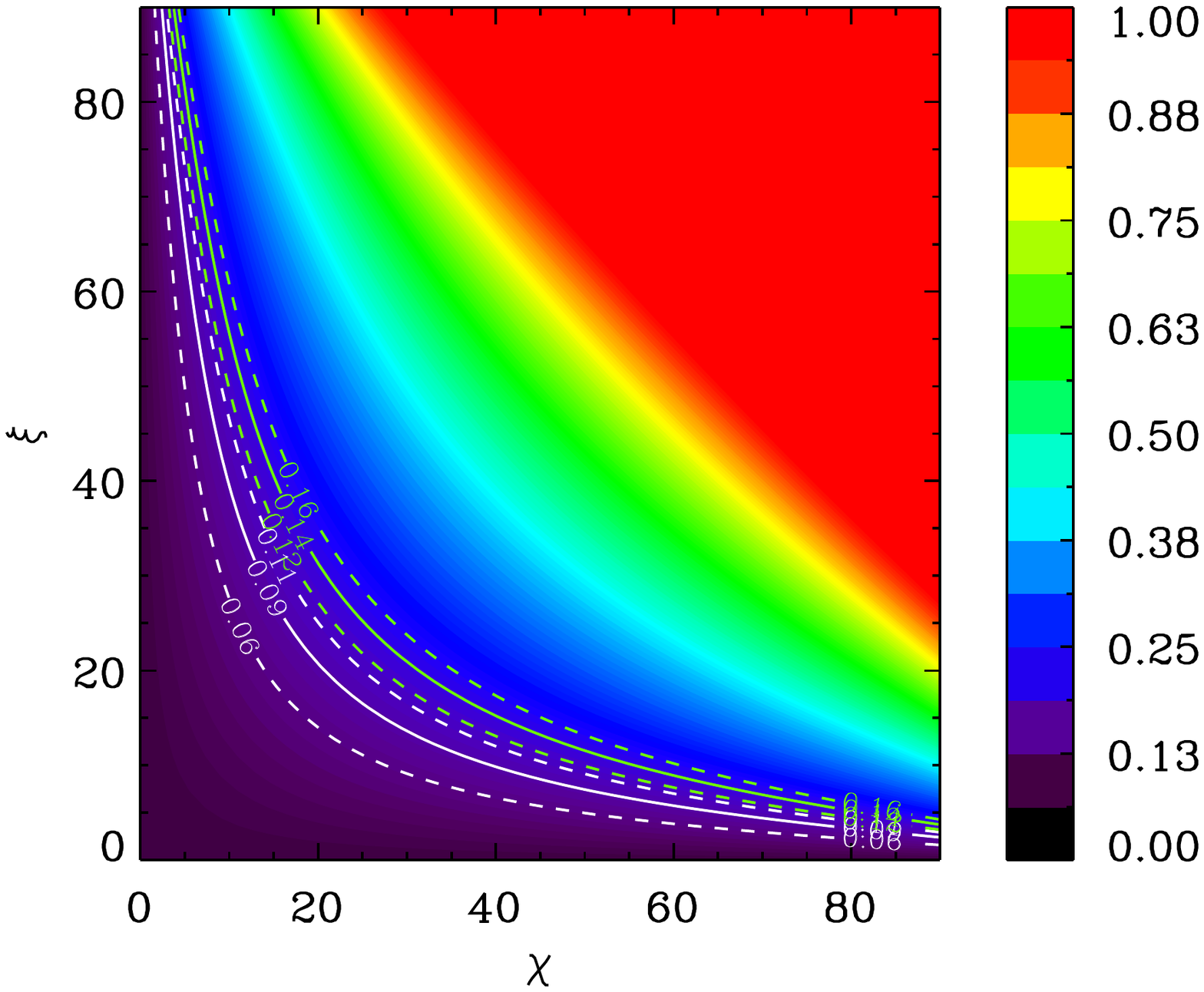}
    \label{fig:model-pf}
    \vspace{-0.5cm}
     \caption{The computed pulsed fraction as a function of the two geometrical angles $\chi$ and $\xi$. The white and green contours correspond to the observed PF for the 2020 October 1 ($9\%$) and of the 2020 May 13 ($14\%$) observation, respectively, together with the associated $1\,\sigma$ errors (dashed lines); see text for details.}
\end{figure}

Pulsations below $1$\,keV were detected in both \xmm\ pointings with a pulsed fraction of $\sim 14\%$ on 2020 May 13 and $\sim 9\%$ on October 1 (this value is however compatible with zero at the 3$\sigma$ level; Figure\,\ref{fig:foldxmm}), and the unabsorbed BB flux decreased of about $20\%$ between the two epochs (Table\,\ref{tab:spec_ave}). The low measured pulsed fraction is consistent with the (nearly) constant values of the BB parameters over the pulse phase, as derived from the phase-resolved spectroscopy (Section\,\ref{sec:phase_res} and Table\,\ref{tab:spec_pps}). 
To gain some insight on the source geometry and on its evolution over the outburst decay, we introduce a simple model according to which thermal photons are produced by a circular cap on the star surface heated at the outburst onset. We assume that the cap is at uniform temperature, as suggested by the lack of multiple BB components in the observed spectrum  \citep[this is at variance with, e.g., the case of XTE\,J1810$-$197,][]{2021MNRAS.504.5244B}. The cap properties are fixed by the measured blackbody temperature ($kT_\mathrm{c}=0.45$\,keV) and radius ($R_\mathrm{BB}=1.6$\,km which results in a semi-aperture $\theta_\mathrm{c}\sim 7\deg$ for $R_\mathrm{NS}=13$ km); these values are representative of both  the \xmm\ observations of 2020 May 13 and October 1 since they do not change significantly between the two epochs. We computed the pulse profiles of the thermal component, as seen by an observer at infinity, as a function of the two geometrical angles $\chi$ and $\xi$ which measure the inclination of the line-of-sight (LOS) and of the cap axis with respect to the rotation axis, respectively. General-relativistic effects are taken into account \citep[see][for details]{2013ApJ...768..147T}\footnote{Interstellar absorption and the detector response function were not accounted for. However, for the particular case we are dealing with (a constant temperature blackbody which changes in phase because of the varying visible area), the pulse profile is independent on both effects and the pulsed fraction is independent on the blackbody temperature.}. Results for the pulsed fraction are shown in Figure\,\ref{fig:model-pf} where the green/white, labeled contour marks the value of the pulsed fraction derived in the XMM observation ID 0871190201 (2020 May 13) and ID 0871191301 (2020 Oct 1), $14\%$ and $9\%$ respectively; the dashed contours are drawn in correspondence to $1\sigma$ errors. Results are not particularly constraining for the source geometry. However, we note that despite the fact that no significant changes in the emission properties of the hot spot were detected, the two values of the PF do not appear to be consistent, at least within $1\sigma$ uncertainties. While this can simply reflect measurement errors, taken face value it may suggest that the hot spot (slightly) changed its position on the surface without sensible variations in size and temperature.

During the X-ray monitoring campaign, the SRT observed \src\ twice, on 2020 October 9 and 10, after the detection of three additional radio bursts by CHIME on October 8 \citep{good20}. Moreover, on October 9, the FAST telescope detected multiple radio pulses with fluence up to 40\,mJy\,ms and pulsed radio emission at a period of $\sim$3.24\,s. During the SRT observations, we did not detect either pulsed emission or radio bursts, setting an upper limit on the flux density for the former of 0.1\,mJy and on the fluence for the latter of 800\,mJy\,ms. Furthermore, a dedicated multi-frequency campaign was initiated with multiple radio facilities after the 2020 April FRB-like radio burst \citep{bailes21} without any successful detections. This phenomenology indicates that \src\ can swing between a radio-loud and a radio-quiet states, although the connection with the X-ray activity currently remains not well understood and will need to be investigated with more coordinated radio and X-ray observations.

\begin{table}
\caption{Log of X-ray bursts detected in the \nicer/XTI and \swift/XRT data sets.\label{tab:new_bursts}}
\begin{tabular}{ccc}
\hline
\hline
Instrument & Obs.ID$^a$ & Burst epoch \\
 & & YYYY-MM-DD hh:mm:ss (TDB) \\ 
\hline
\nicer/XTI       & 3655010201 \#1 & 2020-05-18 07:19:55 \\
                 &  \#2           &  13:39:39           \\
\nicer/XTI       & 3020560106 \#1 & 2020-05-20 14:06:36 \\ 
\nicer/XTI       & 3020560108 \#1 & 2020-05-23 01:52:53 \\
                 &  \#2           & 05:01:59     \\  
\nicer/XTI       & 3655010301 \#1 & 2020-06-17 22:02:05 \\
\nicer/XTI       & 3655010302 \#1 & 2020-06-18 13:20:40   \\
\swift/XRT (WT) & 00033349076 \#1 & 2020-07-24 00:21:13 \\
                & \#2             &  01:42:58  \\
                & \#3             &  01:48:38  \\
\swift/XRT (WT) & 00033349084 \#1 & 2020-09-10 23:56:02 \\
                & \#2             & 23:59:28  \\
                & \#3             & 00:04:29  \\
\swift/XRT (WT) & 00033349085 \#1 & 2020-09-11 09:57:07 \\
                & \#2             & 10:03:43  \\
                & \#3             & 11:18:33 \\
\swift/XRT (WT) & 00033349086 \#1 & 2020-09-17 17:13:47 \\
                & \#2             & 18:51:31 \\
                & \#3             & 18:55:54 \\ 
\swift/XRT (WT) & 00033349087 \#1 & 2020-09-18 00:59:54  \\
                & \#2             & 01:03:55  \\
                & \#3             & 02:36:07 \\
\swift/XRT (WT) & 00033349088 \#1 & 2020-09-19 21:25:43  \\
                & \#2             & 22:59:11  \\
                & \#3             & 23:09:29  \\
                & \#4             & 23:11:05  \\
\swift/XRT (WT) & 00033349090 \#1 & 2020-10-09 22:43:38  \\
                & \#2             & 22:43:47  \\
	            & \#3             & 22:44:30  \\ 
	            & \#4             & 22:44:55  \\
	            & \#5             & 22:45:51  \\
                & \#6             & 22:46:13  \\ 
	            & \#7             & 22:47:29  \\
	            & \#8             & 22:47:43  \\
	            & \#9             & 22:47:58  \\
	            & \#10            & 22:48:08  \\ 
	            & \#11            & 22:48:19  \\   
	            & \#12            & 22:48:24  \\
	            & \#13            & 22:48:31 \\
	            & \#14            & 22:49:16  \\
	            & \#15            & 22:50:10 \\
             	& \#16            & 22:50:12 \\
	            & \#17            & 22:50:17 \\
	            & \#18            & 22:50:46  \\
	            & \#19            & 22:50:47  \\
	            & \#20            & 22:51:29  \\
            	& \#21            & 22:52:03  \\
	            & \#22            & 22:52:43  \\
	            & \#23            & 22:53:39  \\
	            & \#24            & 22:53:44  \\ 
\swift/XRT (WT) & 00033349092 \#1 & 2020-10-11 17:53:53  \\
                & \#2             & 18:04:27  \\
                
\hline
\end{tabular}
\begin{list}{}{}
\item[$^a$]The notation \#N corresponds to the burst number in a given observation.
\end{list}
\end{table}

\section{Acknowledgments}
AB thanks George Younes for useful discussions and suggestions. We thank the referee, T. G\"uver, for his helpful comments. This research is based on observations with \cxo\ (NASA), \nicer\ (NASA), \nustar\ (CaltTech/NASA/JPL), \swift\ (NASA/ASI/UK), \xmm\ (ESA/NASA) and the Sardinia Radio Telescope (SRT). We thank N. Schartel for approving a Target of Opportunity observation with \xmm\ in the Director’s Discretionary Time, and the \xmm\ SOC for carrying out the observation. We also thank the \nicer, \nustar\ and \swift\ teams for promptly scheduling our observations as well as the CHIME/FRB collaboration and the \nicer\ team for asking part of the observations presented in this work. The SRT is funded by the Department of University and Research (MIUR), ASI, and the Autonomous Region of Sardinia (RAS) and is operated as National Facility by the National Institute for Astrophysics (INAF).
A.B. and F.C.Z. are supported by Juan de la Cierva fellowships. A.B., F.C.Z., and N.R. are supported by grants SGR2017-1383, PGC2018-095512-BI00, and the ERC Consolidator grant ``MAGNESIA" (No. 817661). G.L.I., S.M., R.T. and A.T. acknowledge financial support from the Italian MUR through grant PRIN 2017LJ39LM, ``UNIAM". G.L.I. also acknowledges funding from ASI-INAF agreements I/037/12/0 and 2017-14-H.O. This work was partially supported by the program Unidad de Excelencia María de Maeztu CEX2020-001058-M. We also acknowledge support from the PHAROS COST Action (CA16214).
This research has made use of the following software: CIAO (v4.12; \citealt{fruscione06}, MARX (v5.5.1), SAS (v19; \citealt{gabriel04}); HEASoft (v6.28), FTOOLS (v6.28; \citealt{blackburn95}), XSPEC (v12.11.0h; \citealt{arnaud96}), NICERDAS (v7a), NuSTARDAS (v2.0.0), MATPLOTLIB (v3.2.1; \citealt{hunter07}), NUMPY (v1.18.4; \citealt{harris20}).
 
\section{Data availability}

Radio data can be provided by the authors upon reasonable request. Data from \nicer, \swift, \xmm, \cxo\ and \nustar\ observations are publicly available.

\bibliographystyle{mnras}
\bibliography{biblio}


\appendix

\section{\nicer\ bursts: fluence and duration}

In Table~\ref{tab:nicer_bursts}, we list the fluence and durations for the bursts detected in the \nicer/XTI light curves. Owing to uncertainties related to the detector saturation limits, we do not report these quantities for the bursts identified in the \swift/XRT data sets. 

In the table, the fluence refers to the 0.3--10\,keV range and the duration has to be considered as an approximate value. We estimated it by summing the 15.625-ms time bins showing enhanced emission for the structured bursts, and by setting it equal to the coarser time resolution at which the burst is detected in all the other cases.

\begin{table*}
\caption{Log of X-ray bursts detected in the \nicer/XTI light curves. 
\label{tab:nicer_bursts}}
\begin{tabular}{ccccc}
\hline
\hline
Obs.ID & Burst epoch & Fluence & Duration \\
 & YYYY-MM-DD hh:mm:ss (TDB) & (counts) & (ms) \\ 
\hline
3655010201 \#1 & 2020-05-18 07:19:55  & 11 & 62.5 \\
                   \#2           &  13:39:39            & 15 & 62.5 \\
3020560106 \#1 & 2020-05-20 14:06:36  & 7  & 62.5 \\ 
3020560108 \#1 & 2020-05-23 01:52:53  & 6  & 31.25 \\
                   \#2           & 05:01:59     & 7  & 62.5 \\  
3655010301 \#1 & 2020-06-17 22:02:05 & 47 & 62.5 \\
3655010302 \#1 & 2020-06-18 13:20:40 & 6 & 62.5  \\
\hline
\end{tabular}
\end{table*}

\begin{figure*}
    \centering
    \vspace{-0.3cm}
    \includegraphics[width=1.8\columnwidth]{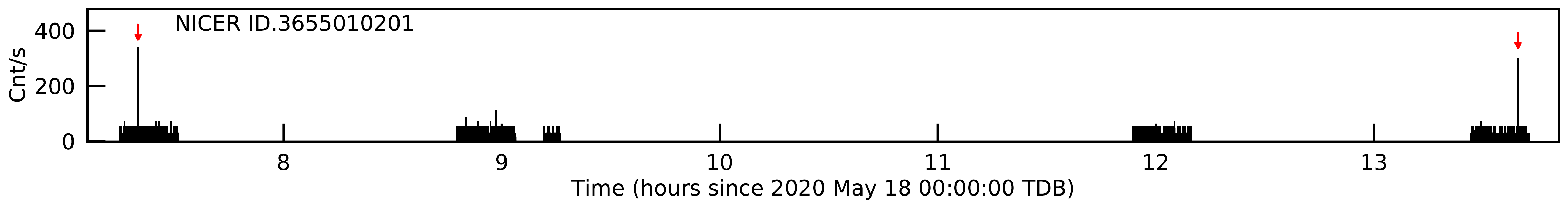}
    \includegraphics[width=1.8\columnwidth]{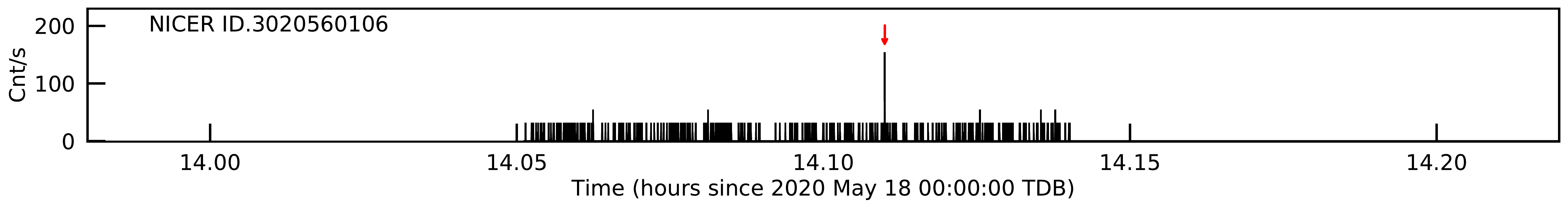}
    \includegraphics[width=1.8\columnwidth]{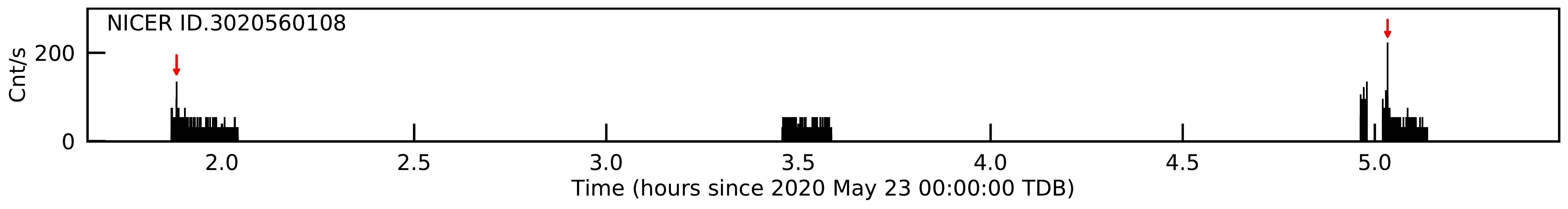}
    \includegraphics[width=1.8\columnwidth]{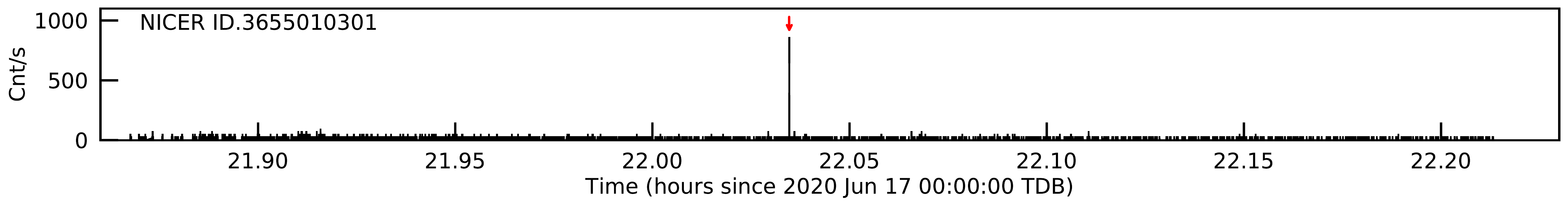}
    \includegraphics[width=1.8\columnwidth]{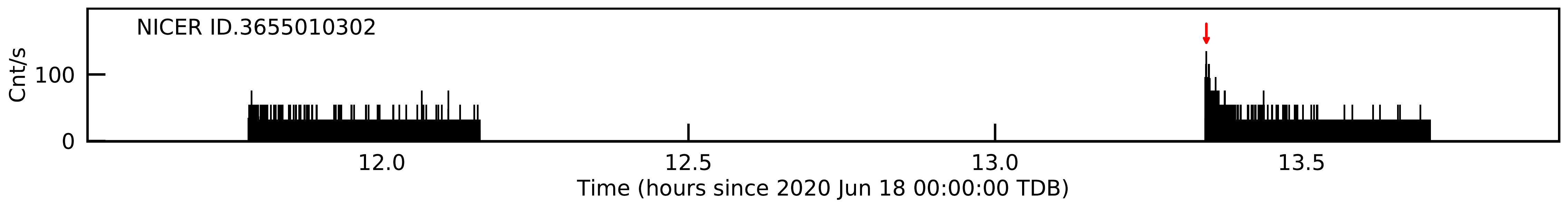}
    \includegraphics[width=1.8\columnwidth]{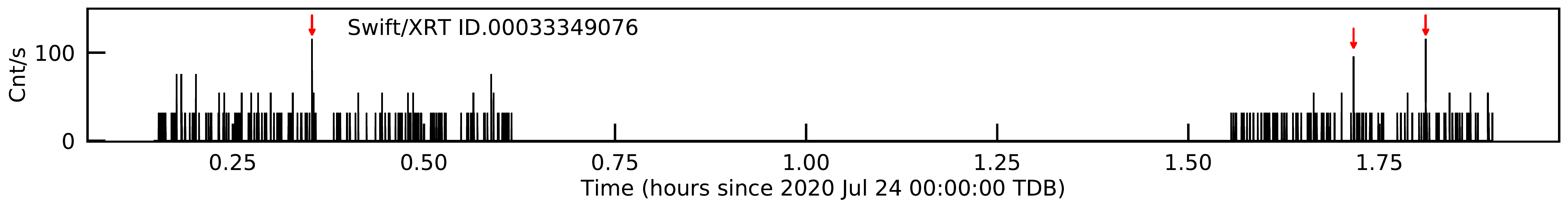}
    \includegraphics[width=1.8\columnwidth]{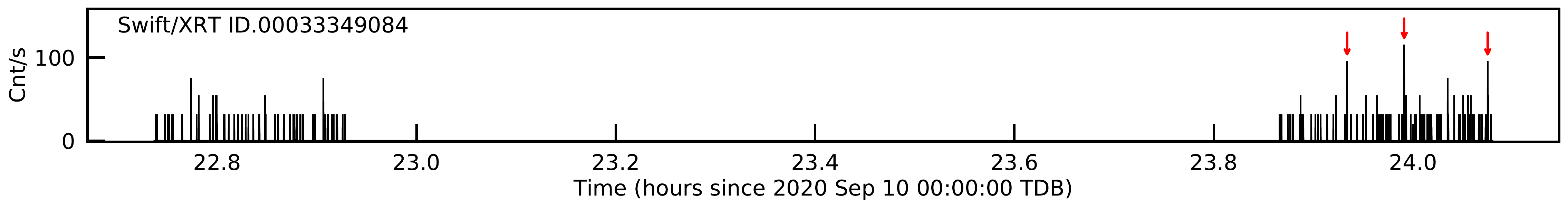}
    \includegraphics[width=1.8\columnwidth]{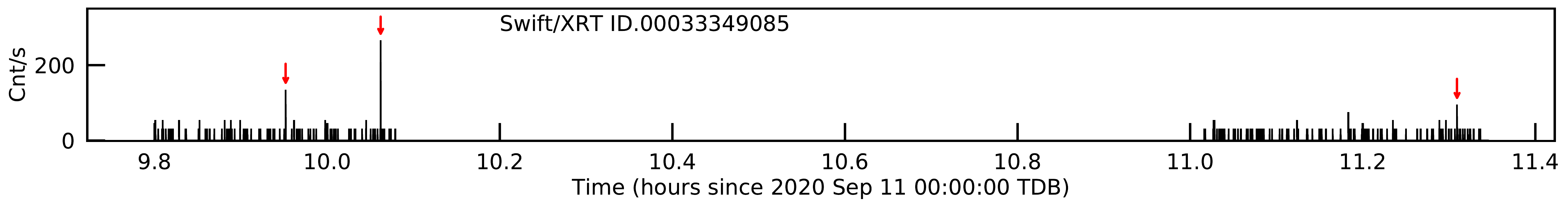}
    \includegraphics[width=1.8\columnwidth]{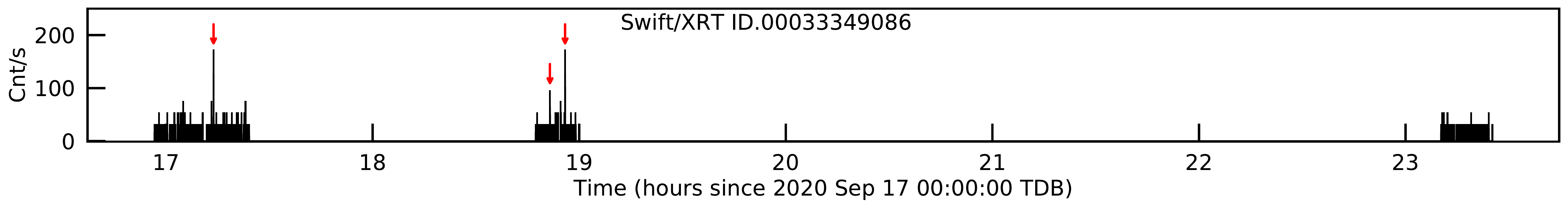}
    \caption{Light curves of \src\ extracted from the \swift/XRT (0.3--10\,keV), \xmm\ (0.3--10\,keV) and \nicer\ (0.3--10\,keV) data in which we detected bursts. All bursts are marked by arrows. The light curves were binned at 62.5\,ms in all cases, except for the \xmm\ data sets where we use a time bin of 73.36\,ms.}
    \label{fig:bursts}
\end{figure*}

\begin{figure*}
    \centering
    \vspace{-0.3cm}
    \includegraphics[width=1.8\columnwidth]{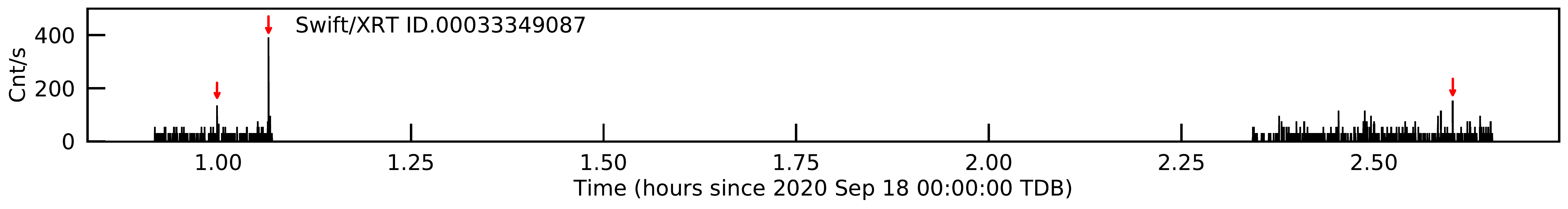}
    \includegraphics[width=1.8\columnwidth]{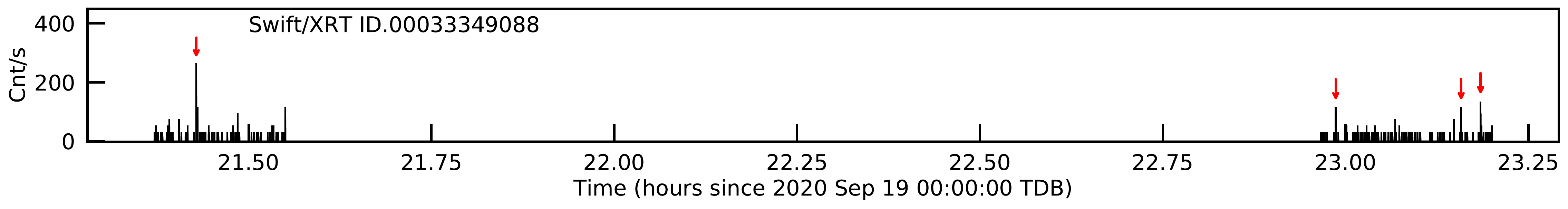}
    \includegraphics[width=1.8\columnwidth]{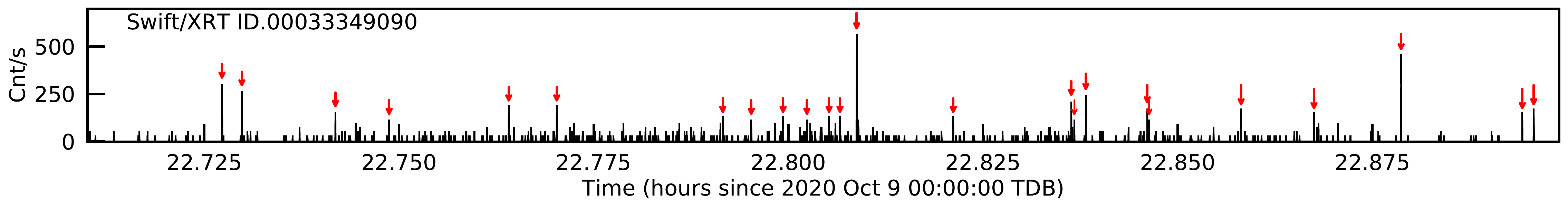}
    \includegraphics[width=1.8\columnwidth]{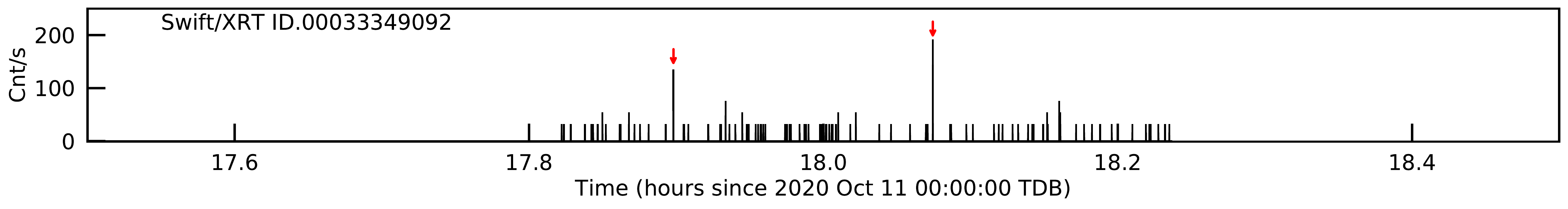}
    \contcaption{}
    \label{fig:bursts}
\end{figure*}



\bsp	
\label{lastpage}
\end{document}